\documentclass[aps,prd, twocolumn, amsmath,amssymb,showpacs,eqsecnum,nofootinbib]{revtex4}

\usepackage{graphicx}
\usepackage{dcolumn}
\usepackage{bm}

\begin{document}

\title{Hadamard Renormalization of the Stress Energy Tensor in a Spherically Symmetric Black Hole Space-Time with
an Application to Lukewarm Black Holes }
\author{Cormac Breen}
\email{cormac.breen@ucd.ie}
\author{Adrian C. Ottewill}
\email{adrian.ottewill@ucd.ie}
\affiliation{School of Mathematical Sciences and Complex \& Adaptive Systems Laboratory, University College Dublin, Belfield, Dublin 4, Dublin, Ireland}

\date{\today}

\begin{abstract}
We consider a quantum field which is in a Hartle-Hawking state propagating in a spherically symmetric black hole space-time. We calculate the components of the stress tensor, renormalized using the Hadamard form of the Green's function, in the exterior region of this  space-time. We then specialize these results to the case of the `lukewarm'  Riessner-Nordstrom-de Sitter black hole.
\end{abstract}


\maketitle

\section{Introduction}
In a recent paper, \cite{Tmunu} henceforth referred to as Paper~\textrm{I}, we calculated the renormalized expectation value of the stress energy tensor operator $\langle \hat{T}^{\mu}_{~\nu} \rangle$ on the horizons of a spherically symmetric space-time. This calculation was performed using the Hadamard renormalization procedure \cite{Wald,BrownOttewill}, which in our view provides the most direct and logical approach to the
renormalization problem for practical calculations. 

Building on the seminal paper by Howard~\cite{Howard:1985yg}, Anderson, Hiscock and Samuel \cite{Anderson:1994hg} developed a method  to compute $\langle \hat{T}^{\mu}_{~\nu} \rangle$ in a general spherically symmetric space-time. Their method relied on the renormalization 
counterterms, denoted by  $\langle\hat{T}_{\mu}^{~ \nu}\rangle_{DS}$, which were first calculated by Christensen \cite{Christensen:1976vb}. This calculation in turn relied on the DeWitt series representation for the Green's function, which is an asymptotic power series in inverse powers of the mass of the field $m$. This series is ill-defined for the massless case and requires some severe modification in order to be applicable to this case. In paper we use the Hadamard renormalization procedure, which is well-defined for both massive and massless fields,  to develop an alternate approach to that of \cite{Anderson:1994hg}. In doing so we renormalise each term in the definition of the stress tensor independently
making it very much easier to debug numerical calculations.

This paper is organized as follows, in Sec.~\ref{sec:Method} we will outline our new method of constructing the stress tensor. The method  leads to renormalized expressions which are readily amenable to numerical computation. In Sec.~\ref{sec:Num} we will outline the numerical method used to calculate these expressions. The results of this method for the lukewarm case are given in Sec.~\ref{sec:Res}. Finally our conclusions are presented in Sec.~\ref{sec:Conclusions}. Throughout this paper we use the sign conventions of Misner, Thorne and Wheeler~\cite{MTW} and work in units in which $8\pi G=\hbar =c =k_B =1$.

\section{Formal Construction}
\label{sec:Method}
In this section we present a new approach to the construction of  $\langle \hat{T}^{\mu}_{~\nu} \rangle_{ren}$ for a spherically symmetric black hole space-time which has a Euclidean line element of the form
\begin{equation}
\label{lee}
\mathrm{d}s^2=f(r) \mathrm{d} \tau^2 +\frac{1}{f(r)} \mathrm{d} r^2 +r^2 \mathrm{d}\theta^2 + r^2 \sin^2 \theta \mathrm{d}\phi^2.
\end{equation}

The standard approach in the literature is to consider an expression for  $\langle \hat{T}^{\mu}_{~\nu} \rangle_{unren}$  in its entirety, then subtract off the renormalizing counterterms, $\langle \hat{T}^{\mu}_{~\nu} \rangle_{DS}$, as calculated by Christensen \cite{Christensen:1976vb}, en masse to obtain a single expression for $\langle \hat{T}^{\mu}_{~\nu} \rangle_{ren}$. This method was first developed by Candelas and Howard for a conformal scalar field in the Schwarzschild space-time \cite{Howard:1984qp,Howard:1985yg} and then extended to case of scalar fields with arbitrary mass and coupling to the gravitational field, in general spherically symmetric space-times, by Anderson, Hiscock and Samuel \cite{Anderson:1994hg}.

We adopt an alternative approach. This approach begins with the definition of the stress tensor given in Paper~\textrm{I}
\begin{align}
\label{eqn:Trent}
\langle\hat{T}^{\mu\nu}\rangle_{ren}=\frac{1}{8\pi^2}\left(\tau^{\mu\nu}[ W_{A}]+ 2 v_1 g^{\mu\nu}\right) + \mathcal{M}^{\mu \nu},
\end{align}
where $\tau^{\mu\nu}[ W_{A}]$ represents the coincidence limit of the differential operator $\tau^{\mu\nu}$ 
\begin{align}
\label{eqn:tau}
\tau^{\mu \nu} &=(1-2 \xi) g_{\nu'}^{~\nu} \nabla^{\mu}\nabla^{\nu'} +(2 \xi -\tfrac{1}{2}) g^{\mu\nu}g_{\alpha'}^{~\alpha}\nabla^{\alpha'}\nabla_{\alpha}\nonumber  \\   &- 2\xi \nabla^{\mu}\nabla^{\nu} + 2 \xi g^{\mu\nu} \nabla_{\alpha}\nabla^{\alpha}\nonumber  \\&+\xi(R^{\mu \nu} -\tfrac{1}{2} R g^{\mu\nu})-\tfrac{1}{2} m^2 g^{\mu\nu},  
\end{align}
acting on the regular part of the Hadamard form for the Euclidean Green's function $G_E(x,x')$. Here 
\begin{align}
\label{eqn:mcalug}
\mathcal{M}^{\mu \nu}=\frac{m^2}{16 \pi^2} \left\{ \left(\xi -\tfrac{1}{6}\right)\left(R^{\mu \nu} -\tfrac{1}{2} g^{\mu \nu} R\right) -\tfrac{3}{8}m^2 g^{\mu \nu}\right\},
\end{align}
 and $g_{\nu}^{~\alpha'}$ is a bivector of parallel transport, which acts to parallel transport a vector at $x'$ to a vector at $x$.

The main idea of our method is that we consider the contribution from each of the derivative terms and from $G_E(x,x')$ to Eq.~(\ref{eqn:Trent}) individually. This involves calculating the mode sum expression for each of these quantities, as well as the corresponding renormalising subtraction terms. The latter quantity comes from consideration of the singular part of Hadamard form of $G_E$, which we will discuss in Section \ref{sec:sub}, while the former may be obtained from the mode sum expression, given in Eq.~(\ref{eqn:modet}) below, and will be the subject matter of Section \ref{sec:unren}. This approach is, of course, completely equivalent to that of Anderson et al. however, we feels it affords a somewhat clearer and more tractable way of constructing $\langle \hat{T}^{\mu}_{~\nu} \rangle_{ren}$.

 A key issue in renormalisation, in particular when one cannot solve the radial equation in closed form, is finding a way of isolating the $x\to x'$ divergences in the components of $\langle \hat{T}^{\mu}_{~\nu} \rangle_{unren}$ which are to cancel with those contained in the corresponding renormalisation subtraction terms. 
For calculations on the horizons of the space-time, radial separation turns out to be the most convenient choice of point separation. 

 Indeed in paper I we made use of  regularization via radial separation to calculate $\langle \hat{T}^{\mu}_{~\nu} \rangle_{ren}$ for a thermal state, on the horizons of a spherically symmetric space-time. In the region exterior to the black hole, excluding the immediate vicinity of the horizons, temporal separation is the favoured choice as angular separation generally leads to very complicated expressions due to derivatives of the Legendre function. We note here that temporal separation only makes sense in this region, as the Killing vector $\partial/\partial t$ become null on the horizons of the black hole and so temporal sepatation cannot be used to calculate horizon values.
 
We will now briefly outline the strategy employed in the construction of the components of $\langle \hat{T}^{\mu}_{~\nu} \rangle_{ren}$ using temporal point splitting before proceeding to describe the details of the calculation. We follow the spirit of \cite{Anderson:1994hg}, but with the differences in approach outlined previously.\\
Consider one component of  $\langle \hat{T}^{\mu}_{~\nu} \rangle_{ren}$, $[g^{\mu \nu'}G_{E}(x,x')_{;\nu \nu'}]_{ren}$ say (the square brackets denote that the coincidence limit has been taken and we choose not to sum over $\nu'$), this is formally given by (with $\epsilon(=t-t') \to 0$) 
\begin{align}
&[g^{\mu \nu'}G_{E}(x,x')_{;\nu \nu'}]_{ren}=
\nonumber\\
 &\lim_{\epsilon \to 0} \left[ g^{\mu \nu'}G_E(x,x')_{;\nu \nu'}-g^{\mu \nu'}G_{Esing}(x,x')_{;\nu \nu' }\right],
\end{align}
where $x'{}^p=x^{p} +\epsilon\delta^{p}_{t}$ and $g^{\mu \nu'}G_{E sing}(x,x')_{;\nu \nu'}$ represents the renormalizing subtraction terms. Note that here we follow the conventions of \cite{Anderson:1994hg}, by performing all our derivatives in Lorentzian space, i.e. the derivatives in Eqn~(\ref{eqn:Trent}) are taken with respect to $x=(t,r,\theta,\phi)$.
In the case of temporal point separation the divergences of the Green's function manifest themselves as divergent mode sums over $n$.
In order to take this coincidence limit, we write the $\epsilon \to 0$ divergences in $g^{\mu \nu'}G_{E sing}(x,x')_{;\mu \nu'}$ as divergent sums over $n$, and then bring these sums inside the mode sum over $n$.  \\
We are then in a position to take the coincidence limit yielding a finite expression which takes the form
\begin{align}
[g^{\mu \nu'}G_{E}(x,x')_{;\nu \nu' }]_{ren}&=[g^{\mu \nu'}G_{E}(x,x')_{;\nu \nu' }]_{numeric}
\nonumber\\
&+[g^{\mu \nu'}G_{E}(x,x')_{;\nu \nu' }]_{analytic},
\end{align}
where $[g^{\mu \nu'}G_{E}(x,x')_{;\nu \nu' }]_{numeric}$ consists of the coincidence limit of the mode sum expression minus the divergent sums over $n$ contained in $g^{\mu \nu'}G_{E sing}(x,x')_{;\nu \nu'}$, and $[g^{\mu \nu'}G_{E}(x,x')_{;\nu \nu' }]_{analytic}$ contains the finite remainder terms from $g^{\mu \nu'}G_{E sing}(x,x')_{;\nu \nu'}$. Note that we are free to take the coincidence limit  in any direction we choose as $W$ is smooth.

\subsection{Unrenormalized Expressions}
\label{sec:unren}
We begin with the mode sum expression for the unrenormalized Euclidean Green's function for a thermal state, with temperature $T=\kappa/2 \pi$, in a spherically symmetric space-time:
\begin{align}
\label{eqn:modet}
G_E(x,x')&=\sum_{n=0}^{\infty} F(n)\cos(n \kappa\left(\epsilon_{\tau})\right)\nonumber\\
&\times\sum_{l=0}^{\infty}(2l+1)P_l(\cos\gamma) p_{nl}(r_<)q_{nl}(r_>),
\end{align}
where $\epsilon_{\tau}=\tau-\tau'$,$r_<=$min$(r,r')$, $r_>=$max$(r,r')$, $F(0)= \kappa/8 \pi^2$ and $F(n)=\kappa/4 \pi^2$, $n>0$. Henceforth we choose to set $r_>=r$. $p_{nl}$ and $q_{nl}$ are the independent solutions to the homogeneous version of this equation, 
    \begin{align}
\label{eqn:mode}
 &\frac{1}{r^2}\frac{d}{dr} \left(r^2 f\frac{d \chi}{dr}\right)-\bigg(\frac{n^2 \kappa_0^2}{f} +\frac{l(l+1)}{r^2} +m^2 +\xi R\bigg) \chi\nonumber\\
& =0
\end{align}
with $p_{nl}$ defined to be the solution which is regular on the lower limit of the region under consideration, while $q_{nl}$ is regular at the upper limit of the region.  $C_{nl}$ is fixed by the Wronskian condition
\begin{align}
C_{nl}\left[p_{nl} \frac{dq_{nl}}{dr}-q_{nl} \frac{dp_{nl}}{dr}\right]=- \frac{1}{r^2 f}.
\end{align}\\
Using this expression we will derive a formal expression $\langle \hat{T}^{\mu}_{~\nu} \rangle_{ren}$ for the Hartle Hawking state in a spherically symmetric space-time.

The first step in our computation is to calculate the required mode sum expressions constructed by action of the appropriate covariant differentiations on Eq. (\ref{eqn:modet}). Firstly considering two derivatives taken at $x$, since $g^{\mu\nu}$ is diagonal it turns out that we only need to calculate derivatives of the form $g^{\mu \mu}G_{E}(x,x')_{;\mu \mu}$ (here $\mu$ is not summed over). For the case of a derivative taken at either point,  $g^{\mu \nu'}G_{E}(x,x')_{;\mu \nu'}$ , the situation is not so straightforward. To see this we consider the world function, $\sigma$ (defined as half the square of the geodesic distance), for temporal splitting, which can be written as an expansion in $\epsilon$ (see Section \ref{sec:sub} for details):
\begin{align}
\sigma=-\frac{1}{2} \epsilon ^2 f-\frac{1}{96} \epsilon
   ^4 \left(f f'^2\right)+\epsilon ^6
   \left(-\frac{f f'^4}{11520}-\frac{f^2
   f'^2 f''}{1920}\right)\nonumber\\
   +O\left(\epsilon
   ^7\right).
\end{align}
Inspection of this expression shows that even though the points are split in time, $\sigma$ has components in the $(t,r)$ plane, due to the fact that the coefficients of the expansion depend on $r$; hence we will have cross components of $g^{\mu\nu'}$ in this plane. The components of the bivectors of parallel transport are readily calculated from the defining equation 
\begin{equation}
\sigma^{; \alpha'}g_{a b' ; \alpha'}=0; \quad [g_{ab'}] =g_{ab},
\end{equation}
and using the non-vanishing components of the connection
\begin{align}
\label{eqn:christ} 
&\Gamma^{r}_{r r}=-\frac{f'(r)}{2 f(r)} ; \quad \Gamma^{r}_{\theta \theta}= -r f(r); \quad \Gamma^{r}_{\phi \phi}= -r f(r) \sin^2\theta;\nonumber\\
&\Gamma^{r}_{tt}=\frac{f(r) f'(r)}{2}
\Gamma^{\theta}_{r \theta}=\Gamma^{\theta}_{\theta r}=\frac{1}{r};\quad
 \Gamma^{\theta}_{\phi \phi}=-\sin\theta \cos\theta;\nonumber\\
 & \Gamma^{\phi}_{r \phi}= \Gamma^{\phi}_{ \phi r}=\frac{1}{r}; \quad  \Gamma^{\phi}_{\theta \phi}= \Gamma^{\phi}_{ \phi \theta} =\cot \theta;
 \Gamma^{t}_{t r}=\Gamma^{t}_{ r t}=\frac{f'(r)}{2 f(r)}.
\end{align}
These bivectors are then given as expansions in $\epsilon\equiv t-t'$ \cite{Anderson:1994hg}:
\begin{align}
&g^{t t'}=-\frac{1}{f}-\frac{\epsilon ^2 f'^2}{8
   f}+\epsilon ^4 \left(-\frac{f'^4}{384
   f}-\frac{1}{96} f'^2
   f''\right)+O\left(\epsilon ^5\right)\nonumber\\
  & g^{tr'}=-g^{rt'}=-\frac{1}{2} \epsilon  f'+\epsilon ^3
   \left(-\frac{1}{48} f'^3-\frac{1}{48} f
   f' f''\right)+O\left(\epsilon ^4\right)\nonumber\\
  & g^{rr'}=f+\frac{1}{8} \epsilon ^2 f
   f'^2+\frac{1}{384} \epsilon ^4 \left(f
   f'^4+4 f^2 f'^2
   f''\right)+O\left(\epsilon ^5\right)\nonumber\\
   &g^{\theta\theta'}=\frac{1}{r^2};\quad   g^{\phi\phi'}=\frac{1}{r^2\sin^2(\theta)}.
\end{align}
Hence $g^{\mu \nu'}G_{E}(x,x')_{;\mu \nu' }$ possesses two off diagonal components, namely $g^{tr'}G_{E}(x,x')_{;tr'}$ and $g^{rt'}G_{E}(x,x')_{;rt' }$.

Armed with these expressions we may now proceed to calculate the unrenormalized expressions for the required derivatives of the Green's function in the partial coincidence limit $(r\to r',\theta\to \theta',\phi \to \phi')$. We choose to denote a bitensor $A(x,x')$ evaluated in this partial coincidence limit by $\{A\}$ and to drop the $(x,x')$ notation for convenience. These expressions are straightforward to derive and so we will just list the final results:

\begin{align}
\{G_E\}=\sum_{n=0}^{\infty}F(n) \cos(n \kappa \epsilon_\tau)\sum_{l=0}^{\infty}(2l+1)p_{nl}(r)q_{nl}(r),
\end{align}
\begin{align}
\{g^{tt}G_{E;t t}\}&=\sum_{n=0}^{\infty} F(n)\cos(n \kappa \epsilon_\tau)\sum_{l=0}^{\infty}(2l+1)\nonumber\\
&\times\left(\frac{ f'}{2}p_{nl}(r)\frac{dq_{nl}(r)}{dr}-\frac{n^2 \kappa^2}{f}p_{nl}(r)q_{nl}(r)\right),
\end{align}
\begin{align}
&\{g^{tt'}G_{E;t t'}\}=\nonumber\\
&-g^{tt'}\sum_{n=0}^{\infty} F(n)n^2 \kappa^2 \cos(n \kappa \epsilon_\tau)\sum_{l=0}^{\infty}(2l+1)p_{nl}(r)q_{nl}(r),
\end{align}
\begin{align}
&\{g^{tr'}G_{E;t r'}\}=\nonumber\\
&-i g^{tr'}\sum_{n=0}^{\infty} F(n)n \kappa \sin(n \kappa \epsilon_\tau)\sum_{l=0}^{\infty}(2l+1)\frac{dp_{nl}(r)}{dr}q_{nl}(r),
\end{align}
\begin{align}
&\{g^{rt'}G_{E;rt'}\}=\nonumber\\
&i g^{rt'}\sum_{n=0}^{\infty} F(n)n \kappa \sin(n \kappa \epsilon_\tau)\sum_{l=0}^{\infty}(2l+1)p_{nl}(r)\frac{dq_{nl}(r)}{dr},
\end{align}
\begin{align}
&\{g^{rr'}G_{E;rr'}\}=\nonumber\\
&g^{rr'}\left(\sum_{n=0}^{\infty} F(n) \cos(n \kappa \epsilon_\tau)\sum_{l=0}^{\infty}(2l+1)\frac{d p_{nl}(r)}{d r}\frac{d q_{nl}(r)}{dr}\right),
\end{align}
\begin{align}
\label{eqn:Gththunrent}
&\{g^{\theta\theta}G_{E;\theta\theta}\}=\{g^{\phi\phi}G_{E;\phi\phi}\}=\nonumber\\
&\sum_{n=0}^{\infty}F(n) \cos(n \kappa \epsilon_\tau)\sum_{l=0}^{\infty}(2l+1)
\left(\frac{ f}{r}p_{nl}(r)\frac{dq_{nl}(r)}{dr}\right.\nonumber\\
&\left.-\frac{ l(l+1)}{2r^2}p_{nl}(r)q_{nl}(r)\right),
\end{align}
\begin{align}
&\{g^{\theta\theta'}G_{E;\theta\theta'}\}= \{g^{\phi\phi'}G_{E;\phi\phi'}\}=\nonumber\\
&\sum_{n=0}^{\infty}F(n) \cos(n \kappa \epsilon_\tau)\sum_{l=0}^{\infty}(2l+1
\left(\frac{ l(l+1)}{2r^2}p_{nl}(r)q_{nl}(r)\right).
\end{align}
Here we have made use of the relation $\partial / \partial t=i \partial/ \partial\tau $. Note that we have not calculated a mode sum expression for $\{g^{rr}G_{E;rr}\}$. This is not required, since as will be seen in Section~\ref{sec:renormal},  $[g^{rr}G_{E;rr}]_{ren}$ may be written in terms of the other renormalized derivatives of the Green's function.

It is well known that all of the sums over $l$ above are divergent \cite{Howard:1984qp}; this has to be a non-physical divergence as the points are still separated. This superfluous divergence is just a reflection of the distributional nature of the Green's function. For our case, with temporal separation, it may be removed by subtracting off appropriate multiples of $\delta(\tau -\tau')$  which vanish when the points are separated. These subtraction terms are easily calculated for each of the above sums, and will be detailed in Section \ref{sec:Num}. For the sake of simplicity,  we will not explicitly incorporate them for the renormalization procedure at this stage; instead when we write down a mode sum over $l$, the subtraction term is included implicitly.

\subsection{Renormalization Subtraction Terms}
\label{sec:sub}
As was discussed earlier, the renormalisation subtraction terms for a particular component are obtained by the action of the corresponding differential operator on the singular part of the Hadamard form of $G_{E}$, namely
\begin{equation}
\label{eqn:Gsing}
G_{sing} (x ,x') = \frac{\Delta^{1/2}}{\sigma} + V \log(\lambda \sigma).
\end{equation}
Here $\lambda$ is a constant which is required to ensure that the argument of the logarithm is dimensionless. Following the convention of Christensen \cite{Christensen:1976vb}, we let $\lambda= e^{2\gamma} \mu^2/2$, where $\mu=m$ for a massive field but is arbitrary when the field is massless \cite{Anderson:1994hg}. Following \cite{Anderson:1994hg}, we choose to set $\mu=1$ for massless fields.
We make use of a method due to Ottewill and Wardell \cite{Barry:non-geo} to obtain the desired subtraction terms, which we will now briefly outline.

We begin by expanding $\sigma$ in terms of an expansion in powers of the coordinate separation of $x$ and $x'$, $\Delta x^{\alpha}$:
\begin{align}
\label{eqn:coord}
\sigma=\frac{1}{2} g_{\alpha\beta}\Delta x^{\alpha}\Delta x^{\beta} + A_{\alpha\beta\gamma}\Delta x^{\alpha}\Delta x^{\beta}\Delta x^{\gamma}\nonumber\\
 +  B_{\alpha\beta\gamma\delta}\Delta x^{\alpha}\Delta x^{\beta}\Delta x^{\gamma}\Delta x^{\delta}+\dots
\end{align} 
where the coefficients $A_{\alpha\beta\gamma},B_{\alpha\beta\gamma\delta}$ are given by
\begin{align}
A_{abc}&=-\frac{1}{4} g_{(ab,c)};\\
B_{abcd}&=-\frac{1}{3} \bigg( A_{(abc,d)} + g^{\alpha\beta}\bigg(\frac{1}{8} g_{(ab,|\alpha|}A_{|\beta|cd)} \nonumber\\
&+\frac{9}{2}A_{\alpha(ab}A_{|\beta|cd)}\bigg)\bigg).
\end{align}
The higher order terms are easily calculated by hand or by using Mathematica \cite{Mathematica}. 
Here we have followed the standard convention of denoting symmetrization of indices by using brackets (e.g. $(\alpha\beta )$) and exclude indices from symmetrization by surrounding them 
by vertical bars (e.g. ($\alpha 
|\beta |\gamma $)). 

Wardell and Ottewill \cite{BarryWeb} have developed a Mathematica notebook which allows one to expand both $\Delta^{1/2}(x,x')$ and $V(x,x')$ in a coordinate expansion in powers of the coordinate separation of $\Delta x^{\alpha}$, using Eq. (\ref{eqn:coord}),  for arbitrary point splitting to high order. This notebook thus allows one to obtain a series expansion of $G_{sing} (x ,x')$ in powers of $\Delta x^{\alpha}$, up to the required order to capture both the divergence and the finite remainder terms of $G_{sing} (x ,x')$ and its derivatives in the coincidence limit. 

The results of this procedure for temporal point splitting are given in Appendix \ref{ap:Sub}. 
  
  \subsection{Renormalization} 
  \label{sec:renormal} 
  We now proceed to obtain renormalized expressions for each of the derivatives of the Green's function required to construct the renormalized stress tensor.
 As outlined previously, we achieve this by writing the geometric $x \to x'$ divergences in the subtraction terms as divergent mode sums over $n$ and then taking the coincidence limit. To do this we make use of the following identities, which all follow from the first expression, found in \cite{gradriz}:
  \begin{align}
  \label{eqn:divsums}
  \kappa \sum_{n=1}^{\infty} \frac{\cos(n \kappa \epsilon_{\tau})}{n \kappa} &=-\frac{1}{2} \ln(-\kappa^2 \epsilon^2) +O(\epsilon^2);\nonumber\\
    i \kappa \sum_{n=1}^{\infty}\sin(n \kappa \epsilon_{\tau})& =\frac{1}{\epsilon} +O(\epsilon);\nonumber\\
   \kappa \sum_{n=1}^{\infty}n \kappa \cos(n \kappa \epsilon_{\tau})&=\frac{1}{\epsilon^2} -\frac{\kappa^2}{12}+O(\epsilon^2);\nonumber\\
       i\kappa \sum_{n=1}^{\infty}n^2 \kappa^2 \sin(n \kappa \epsilon_{\tau})&=\frac{2}{\epsilon^3}+O(\epsilon^2)\nonumber\\
    \kappa \sum_{n=1}^{\infty}n^3 \kappa^3 \cos(n \kappa \epsilon_{\tau})&=\frac{6}{\epsilon^4}+\frac{\kappa^4}{120}+O(\epsilon^2).
  \end{align}

It is at this stage of the calculation that the main advantage of our method becomes apparent. In previous calculations  \cite{Anderson:1994hg,Howard:1985yg}, these identities were applied to the divergences contained in the Christensen subtraction terms, resulting in a collection of divergent  sums over $n$. These sums then had to be distributed among the relevant mode sum expressions  in a manner which rendered each sum convergent. The order of this distribution, however, was far from obvious. In fact, in order to ensure the correct distribution, one had to first insert the WKB approximations to the radial solutions into the mode sum expression, then perform these sums in the large $n$ limit to see which counter-terms were needed for each sum. For our method, however, the order of distribution is immediate, as the correct divergent $n$ sum for each component must come from that particular components renormalizing counter term.

We now employ this method to obtain a finite expression for each of the individual derivatives of the Green's function required to construct the renormalized stress tensor. These calculations are repetitive, so will show the details of the calculation of one component, $[G_{E}]_{ren}$, and simply list the results for the others. We note here that for simplicity, we choose to perform our calculations for a space-time with a constant Ricci scalar $R$. Therfore we may define an effective field mass $\hat{m}=\sqrt{m^2 +(\xi -1/6)R}$. It is worth noting also that space-times with a cosmological constant that are solutions of Einstein's vacuum equations, possess a constant Ricci scalar. 

  Using the identities in Eq. (\ref{eqn:divsums}) we may write the subtraction terms $G_{Esing}$ in the form
  \begin{align}
 8\pi^2\{G_{Esing}\}=-\kappa\sum_{n=1}^{\infty}\cos(n\kappa\epsilon_{\tau})\left(\frac{2}{f} n \kappa +\frac{\hat{m}^2 }{n \kappa}\right) -\frac{\kappa^2}{6 f} \nonumber\\
  +\frac{\hat{m}^2}{2} \ln\left(\frac{\lambda f}{2 \kappa^2}\right)-\frac{f''}{12}
   +\frac{f'^2}{24 f} -\frac{f'}{6r}.
   \end{align}   
     Hence we can write down an expression for $[G_{E}]_{ren}$ for which the coincidence limit can now be readily taken:
     \begin{align}
    &[G_{E}]_{ren}=\lim_{\epsilon_{\tau} \to 0}\left\{\frac{\kappa}{4\pi^2} \sum_{n=1}^{\infty}\cos(n\kappa\epsilon_{\tau})\right.\nonumber\\
    &\left.\times\left(\sum^{\infty}_{l=0}(2l+1)p_{nl}q_{nl}+\frac{1}{f} n \kappa + \frac{\hat{m}^2 \kappa}{2 n \kappa}\right)\right.\nonumber\\
     &  \left.+\frac{\kappa}{8\pi^2}\sum^{\infty}_{l=0}(2l+1)p_{0l}q_{0l}
+\frac{\kappa^2}{48\pi^2 f} -\frac{\hat{m}^2}{16 \pi^2} \ln\left(\frac{\lambda f}{2 \kappa^2}\right) \right.\nonumber\\
     &  \left. +\frac{f''}{96\pi^2} -\frac{f'^2}{192\pi^2f} +\frac{f'}{48\pi^2r}\right\}.
    \end{align}
    We may now simply take the coincidence limit yielding the result
      \begin{align}
 [G_{E}]_{ren}= [G_{E}]_{numeric}+ [G_{E}]_{analytic},
\end{align}
where
         \begin{align}
 [G_{E}]_{numeric}&=\frac{\kappa}{4\pi^2} \sum_{n=1}^{\infty}\left(\sum^{\infty}_{l=0}(2l+1)p_{nl}q_{nl}+\frac{1}{f} n \kappa +\frac{\hat{m}^2 }{2 n \kappa}\right)\nonumber\\
&+\frac{\kappa}{8\pi^2}\sum^{\infty}_{l=0}(2l+1)p_{0l}q_{0l},\nonumber\\
[G_{E}]_{analytic}&=\frac{\kappa^2}{48\pi^2 f} -\frac{\hat{m}^2}{16 \pi^2} \ln\left(\frac{\lambda f}{2 \kappa^2}\right)+\frac{f''}{96\pi^2}\nonumber\\
& -\frac{f'^2}{192\pi^2f } +\frac{f'}{48\pi^2r}.
\end{align}
Applying this procedure to the other components yields the results:
\begin{widetext}
 \begin{align}
[g^{rt'}G_{E; rt' }]_{ren}=[g^{tr'}G_{E; tr' }]_{ren}=0,
\end{align}      
              \begin{align}
&[g^{tt'}G_{E;tt'}]_{numeric}=\frac{\kappa}{4 \pi^2 f}\sum_{n=1}^{\infty} \left[n^2 \kappa^2\sum_{l=0}^{\infty}(2l+1)p_{nl}(r)q_{nl}(r)+ \frac{n^3 \kappa^3}{f} +\frac{m^2}{2}n \kappa-\frac{L_{tt'}}{n \kappa}\right]\nonumber\\
&[g^{tt'}G_{E;tt'}]_{analytic}=-\frac{\kappa^4}{480 \pi^2 f^2} +\frac{\hat{m}^2 \kappa^2}{96 \pi^2f }+\frac{L_{tt'}}{8 \pi^2f } \ln\left(\frac{\lambda f}{2\kappa^2}\right)+\frac{F_{tt'}}{8 \pi^2f },
\end{align}
  \begin{align}
[g^{rr'}G_{E;rr' }]_{numeric}&=\frac{\kappa f }{4 \pi^2 }\sum_{n=1}^{\infty}  \left[\sum_{l=0}^{\infty}(2l+1)\frac{d p_{nl}(r)}{d r}\frac{d q_{nl}(r)}{dr}-\frac{ n^2 \kappa^2}{3 f^3} -\left(\frac{f''}{6f^2} -\frac{ f'^2}{3f^3} +\frac{m^2}{2f^2}\right)n \kappa+\frac{L_{rr'}}{n\kappa}\right]\nonumber\\
&+\frac{\kappa f}{8 \pi^2  }\sum_{l=0}^{\infty}(2l+1)\frac{d p_{0l}(r)}{d r}\frac{d q_{0l}(r)}{dr},\nonumber\\
[g^{rr'}G_{E;rr' }]_{analytic}&=\frac{\kappa^4}{1440\pi^2 f^2} -\frac{\kappa^2}{96 \pi^2}\left(\frac{f''}{3f} -\frac{2 f'^2}{3f^2} +\frac{m^2}{f}\right) -f \frac{L_{rr'}}{8 \pi^2} \ln\left(\frac{\lambda f}{2 \kappa^2}\right) -f \frac{F_{rr'}}{8\pi^2},
\end{align}
 \begin{align}
[g^{\theta\theta'}G_{E;\theta\theta'}]_{numeric}&=\frac{\kappa}{4 \pi^2}\sum_{n=1}^{\infty}\left[\sum_{l=0}^{\infty}(2l+1)\frac{ l(l+1)}{2r^2}p_{nl}(r)q_{nl}(r)-\frac{1}{3 f^2} n^3 \kappa^3 -\frac{Q_{\theta\theta'}}{2}n \kappa +\frac{L_{\theta\theta'}}{n \kappa}\right]
\nonumber\\
&+\frac{\kappa}{16 \pi^2r^2}\sum_{l=0}^{\infty}(2l+1)\ l(l+1)p_{0l}(r)q_{0l}(r),\nonumber\\
[g^{\theta\theta'}G_{E;\theta\theta'}]_{analytic}&=\frac{\kappa^4}{1440 \pi^2 f^2}-\frac{Q_{\theta\theta'}\kappa^2}{96\pi^2} -\frac{L_{\theta\theta'}}{8\pi^2}\ln\left(\frac{\lambda f}{2\kappa^2}\right) -\frac{F_{\theta\theta'}}{8\pi^2},
\end{align}
 \begin{align}
[g^{tt}G_{E;tt}]_{numeric}&=\frac{\kappa f'}{8 \pi^2}\sum_{n=1}^{\infty} \left[\sum_{l=0}^{\infty}(2l+1)p_{nl}(r)\frac{dq_{nl}(r)}{dr}-\frac{f'}{2f^2} n\kappa \right]+\frac{\kappa f'}{16 \pi^2}\sum_{l=0}^{\infty}(2l+1)p_{0l}(r)\frac{dq_{0l}(r)}{dr}-[g^{tt'}G_{E;tt'}]_{numeric},\nonumber\\
[g^{tt}G_{E;tt}]_{analytic}&=  \frac{\kappa^4}{480\pi^2 f^2} -\left(\frac{m^2}{f} +\frac{f'^2}{2f^2}\right)\frac{\kappa^2}{96\pi^2}  -\frac{L_{tt}}{8\pi^2}\ln\left(\frac{\lambda f}{2\kappa^2}\right)   -\frac{F_{tt}}{8\pi^2}.
\end{align}
\begin{align}
&[g^{\theta\theta}G_{E;\theta\theta}]_{numeric}=\frac{2 f}{f' r}\left([g^{tt}G_{E;tt}]_{numeric}+[g^{tt'}G_{E;tt'}]_{numeric}\right)-[g^{\theta\theta'}G_{E;\theta\theta'}]_{numeric},\nonumber\\
&[g^{\theta\theta}G_{E;\theta\theta}]_{analytic}=-\frac{\kappa^4}{1440 \pi^2 f^2} -\frac{Q_{\theta\theta} \kappa^2}{96\pi^2}-\frac{L_{\theta\theta}}{8\pi^2}\ln\left(\frac{\lambda f}{2\kappa^2}\right)-\frac{F_{\theta\theta}}{8\pi^2}.
\end{align}
With
  \begin{align}
  L_{tt'}&= -\frac{f}{1440 r^4}\left\{r^4 f'' \left(f''+60 \hat{m}^2\right)+4 r^2
   f'^2-2 r^3 f' \left(r f'''+2 f''-60
   \hat{m}^2\right)-4 r f \left(2 f'+r \left(r^2
   f''''+3 r f'''-f''\right)\right)\right.\nonumber\\
   &\left.+4
   f^2-4 \left(45 \hat{m}^4 r^4+1\right)\right\},\nonumber\\
 F_{tt'}&=   \frac{1}{2880
   r^4 f}\left\{3 r^4 f'^4-12 r^4 f f'^2
   \left(f''+5 \hat{m}^2\right)+4 f^2 \left(r^2
   \left(3 r^2 f'' \left(f''-30 \hat{m}^2\right)-4
   f'^2\right.\right.\right.\nonumber\\
   &\left.\left.\left.+r f' \left(9 r f'''+28 f''-180
   \hat{m}^2\right)\right)+270 \hat{m}^4 r^4+6\right)+24 r f^3
   \left(2 f'+r \left(r^2 f''''  +3 r
   f'''-f''\right)\right)-24 f^4\right\}.
   \end{align}
     \begin{align}
L_{rr'}&= \frac{1}{1440 r^4 f }\left\{r^4 f'' \left(f''+60 \hat{m}^2\right)+4 r^2
   f'^2-2 r^3 f' \left(r f'''+2 f''-60
   \hat{m}^2\right)+4 r f \left(r \left(r
   f'''+f''\right)-2 f'\right)\right.\nonumber\\
   &\left.+4 f^2-4
   \left(45 \hat{m}^4 r^4+1\right)\right\},\nonumber\\
   F_{rr'}&=\frac{1}{2880 r^4 f^3}\left\{41 r^4 f'^4-4 r^4 f f'^2 \left(26
   f''+15 m^2\right)+4 f^2 \left(10 r^4 f''
   \left(f''+9 m^2\right)+r^3 f' \left(5 r
   f'''-24 f''+60 m^2\right)\right.\right.\nonumber\\
   &\left.\left.-90 m^4
   r^4-2\right)-8 r f^3 \left(26 f'+r \left(7
   r^2 f''''+15 r f'''-25
   f''\right)\right)+8 f^4\right\}.
    \end{align} 
   \begin{align}
     Q_{\theta\theta'} &=  \frac{1}{6 r^2 \epsilon ^2  f^2}\left(-r^2 f'^2+f \left(r^2 f''+2 r f'+6
   \hat{m}^2 r^2+2\right)-2 f^2\right)\nonumber\\
  L_{\theta\theta'} &=  -\frac{1}{1440 r^4}\left\{r^4 f''^2+4 r^2 f'^2-2 r^3 f' \left(r
   f'''+2 f''+60 \hat{m}^2\right)-2 r f \left(4
   f'+r \left(r^2 f''''+2 r f'''-2
   f''+60 \hat{m}^2\right)\right)\right.\nonumber\\
 &  \left.+4 f^2+180 \hat{m}^4
   r^4+120 \hat{m}^2 r^2-4\right\},\nonumber\\
    F_{\theta\theta'} &=\frac{1}{2880 r^4 f^2}\left\{11 r^4 f'^4-8 r^2 f^3 \left(r \left(9
   f'''+r f''''\right)+16 f''\right)-2
   r^2 f f'^2 \left(17 r^2 f''+10 r f'+30
   \hat{m}^2 r^2+10\right)\right.\nonumber\\
&   \left.+4 f^2 \left(90 \hat{m}^4
   r^4-2\right)+8 f^4\right.\nonumber\\
   &\left.+4 f^2 \left(r \left(5 r
   f'' \left(r^2 f''+6 \hat{m}^2 r^2+2\right)-13 r
   f'^2+f' \left(5 r^3 f'''+14 r^2
   f''+120 \hat{m}^2 r^2+20\right)\right)\right)\right\}.
   \end{align}
     \begin{align}
   L_{tt}&= \frac{L_{tt'}}{f},\nonumber\\
     F_{tt}&=   \frac{1}{2880
   r^4 f^2}\left\{-27 r^4 f'^4+12 r^4 f f'^2 \left(4
   f''+25 \hat{m}^2\right)+4 f^2 \left(3 \left(r^4
   f'' \left(f''-30 \hat{m}^2\right)+90 \hat{m}^4
   r^4+2\right) + 26 r^2 f'^2\right.\right.\nonumber\\
 & \left. \left.-2 r^3 f' \left(3 r
   f'''+f''+90 \hat{m}^2\right)\right)+24 r f^3
   \left(2 f'+r \left(r^2 f'''+3 r
   f'''-f''\right)\right)-24 f^4\right\}.
   \end{align}
    \begin{align}
 Q_{\theta\theta}=\frac{f'}{r f} -Q_{\theta\theta'};\quad L_{\theta\theta}=-L_{\theta\theta'},
 \end{align}
 \begin{align}
F_{\theta\theta}&= \frac{1}{2880 r^4 f^2}\left\{-11 r^4 f'^4+2 r^2 f f'^2 \left(17 r^2
   f''-20 r f'+30 \hat{m}^2 r^2+10\right) +4 f^2
   \left(-5 r^2 f'' \left(r^2 f''+6 \hat{m}^2
   r^2+2\right)\right.\right.\nonumber\\
  &\left. \left.+13 r^2 f'^2+r f' \left(-5 r^3
   f'''+16 r^2 f''+60 \hat{m}^2 r^2-20\right)+90
   \hat{m}^4 r^4+2\right)\right.\nonumber\\
 &\left.  +8 r f^3 \left(30 f'+r
   \left(r^2 f''''-6 r f'''-14
   f''\right)\right)-8 f^4\right\}.
\end{align}
The construction of  $[g^{rr}G_{E;rr}]_{ren}$ follows immediately from the other components. To see this we exploit  the fact that $[W]$ (the coincidence limit of the regular part of the Hadamard expansion for $G_{E}$) satisfies the inhomogeneous wave equation \cite{BrownOttewill}
 \begin{align}
 (\Box -m^2 -\xi R)[W] = -6v_1,
 \end{align}
  where for a static spherically symmetric  Ricci-constant space-time $v_1$ is of the form
   \begin{align}
 v_1 &=\tfrac{1}{720} R_{abcd}R^{abcd} -\tfrac{1}{720} R_{ab}R^{ab} +\frac{\hat{m}^4}{8}\nonumber\\
 &=\frac{1}{1440 r^4}\left\{r^4 f''^2-8 f \left(r f'+1\right)+f' \left(8 r-4 r^3
   f''\right)+4 f^2+180 \hat{m}^4 r^4+4\right\}.
 \end{align}
Now since $[g^{rr}G_{E;rr}]_{ren}=[g^{rr}W_{;rr}]$ by definition, we have the result that
\begin{align}
[g^{rr}G_{E;rr}]_{ren}=[g^{rr}G_{E;rr}]_{numeric}+[g^{rr}G_{E;rr}]_{analytic},
\end{align} 
where
\begin{align}
[g^{rr}G_{E;rr}]_{numeric}&=-[g^{tt}G_{E;tt}]_{numeric}-2[g^{\theta\theta}G_{E;\theta\theta}]_{numeric}
 +(m^2+\xi R)[G_{E}]_{numeric}\nonumber\\
[g^{rr}G_{E;rr}]_{analytic}&=-[g^{tt}G_{E;tt}]_{analytic}-2[g^{\theta\theta}G_{E;\theta\theta}]_{analytic}+(m^2+\xi R)[G_{E}]_{analytic}-\frac{3v_1}{4\pi^2}.
\end{align} 
\end{widetext}
The extension to non-Ricci constant space-times is straightforward and results in an extra $1/n\kappa$ contribution to the sum over $n$ from the renormalization counter-terms (proportional to $R'$). We choose not to pursue this extension here as we are interested in solutions to Einstein's equations.

\subsection{Formal Expressions}
 In static, spherically symmetric space-times, states respecting the same symmetries have a stress tenor $\langle \hat{T}^{\mu}_{~\nu}\rangle_{ren}$, which is diagonal. We may now write down formal expressions for the diagonal elements of $\langle \hat{T}^{\mu}_{~\nu}\rangle_{ren}$ in the following manner: 
\begin{align}
\langle \hat{T}^{\mu}_{ \nu} \rangle_{ren}=\langle \hat{T}^{\mu}_{ \nu} \rangle_{numeric}+\langle \hat{T}^{\mu}_{ \nu} \rangle_{analytic}
\end{align}
where (not summing over $\nu$)
\begin{align}
&\langle \hat{T}^{\nu}_{~\nu}  \rangle_{numeric}=2(\tfrac{1}{2} -\xi)g^{\nu\nu'}[G_{E;\nu\nu'}]_{numeric} \nonumber\\
&+(2\xi -\tfrac{1}{2})[g^{\alpha\alpha'}G_{E;\alpha \alpha '}]_{numeric}-2\xi [g^{\nu\nu}G_{E;\nu\nu}]_{numeric}
\nonumber\\
&+2\xi [g^{\alpha \alpha}G_{E;\alpha \alpha}]_{numeric}+\xi(R^{\nu}_{~\nu} -\tfrac{1}{2}R)[G_{E}]_{numeric}\nonumber\\
&-\frac{m^2}{2}[G_{E}]_{numeric},
\end{align}
and
\begin{align}
\label{eqn:Tanal}
&\langle \hat{T}^{\nu}_{~\nu} \rangle_{analytic}=\nonumber\\
&2(\tfrac{1}{2} -\xi)g^{\nu\nu'}[G_{E;\nu\nu'}]_{analytic}+(2\xi -\tfrac{1}{2})[g^{\alpha\alpha'}G_{E;\alpha \alpha '}]_{analytic}
\nonumber\\
&-2\xi [g^{\nu,\nu}G_{E;\nu,\nu}]_{analytic}+2\xi [g^{ \alpha \alpha}G_{E;\alpha \alpha}]_{analytic}
\nonumber\\
&+\xi(R^{\nu}_{~\nu} -\tfrac{R}{2})[G_{E}]_{analytic}-\frac{m^2}{2}[G_{E}]_{analytic}+\frac{2 v_1}{8 \pi^2}\nonumber\\
& +\mathcal{M}^{\nu}_{~\nu},
\end{align}
with
\begin{align}
\mathcal{M}^{\nu}_{~\nu}=\frac{m^2}{16 \pi^2} \left\{ \left(\xi -\frac{1}{6}\right)\left(R^{\nu}_{~ \nu} -\frac{1}{2} R\right)-\frac{3}{8} m^2\right\}.
\end{align}
The components of the Ricci tensor are given by
\begin{align}
R^{t}_{~t}=R^{r}_{~r}=-\frac{f''}{2} -\frac{f'}{r},\nonumber\\
R^{\theta}_{~\theta}=R^{\phi}_{~\phi}=\frac{1}{r^2} -\frac{f}{r^2} -\frac{f'}{r}.
\end{align}
If we were able to solve the radial equation (\ref{eqn:mode}) in closed form and perform the required mode sums we would now possess everything required to calculate each component of $\langle \hat{T}^{\mu}_{~\nu}  \rangle_{ren}$. Unfortunately, in general, the radial equation must be solved numerically, adding more complexity to the calculation. This will be the subject matter of the Section \ref{sec:Num}

\subsection{Conservation Equations}
\label{sec:conserva}
In order for the expressions we have just obtained for $\langle \hat{T}^{\mu}_{~\nu} \rangle_{ren}$ to be correct they must, by Wald's axioms \cite{Wald}, satisfy the conservation equations, $\nabla_{\mu}\langle \hat{T}^{\mu}_{~\nu} \rangle_{ren}=0 $.
Using the symmetries of the space-time and Eq.~(\ref{eqn:christ}) it is straightforward to show that the only equation which is not identically satisfied is: 
\begin{align*}
\nabla_{\mu}\langle \hat{T}^{\mu}_{~r} \rangle_{ren} &=\langle \hat{T}^{\mu}_{~r}  \rangle_{ren, \mu} + \langle \hat{T}^{\alpha}_{~r} \rangle_{ren}\Gamma^{\mu}_{\alpha \mu} -\langle \hat{T}^{\mu}_{~\alpha} \rangle_{ren}\Gamma^{\alpha}_{r \mu}\nonumber\\
\end{align*}
which, again using Eq. (\ref{eqn:christ}), reduces to
\begin{align}
\label{eqn:conser}
\nabla_{\mu}\langle \hat{T}^{\mu}_{~r} \rangle_{ren}&=\langle \hat{T}^{r}_{~r} \rangle_{ren,r} +\frac{2}{r} \langle \hat{T}^{r}_{~r} \rangle_{ren}+\frac{f'}{2f}(\langle \hat{T}^{r}_{~r} \rangle_{ren}\nonumber\\
&-  \langle \hat{T}^{t}_{~t} \rangle_{ren}) -\frac{2}{r}  \langle \hat{T}^{\theta}_{~\theta} \rangle_{ren}.
\end{align}
At this juncture we are in a position to check that the analytic components of the stress tensor satisfy the above equation. To do this we substitute the relevant expressions obtained for the analytic components obtained in Section \ref{sec:renormal} into Eq. (\ref{eqn:Tanal}) and then in turn substitute these into the conservation equation  (\ref{eqn:conser}).  Due to the algebraic complexity of the expression involved, this procedure is most easily done in a Mathematica  notebook and the result is that the right hand side of Eq. (\ref{eqn:conser}) vanishes. Therefore we may conclude that the analytic contribution to $\langle \hat{T}^{\mu}_{~\nu}  \rangle_{ren}$ is a conserved quantity, i.e.
\begin{align}
\nabla_{\mu}\langle \hat{T}^{\mu}_{~\nu} \rangle_{analytic}=0.
\end{align}

\section{Numerical Calculation for the  Lukewarm Black Hole}
\label{sec:Num} 

In this section we describe the details of the numerical calculations required in order to calculate both 
 $\langle \phi^2\rangle_{ren}$ and $\langle \hat{T}^{\mu}_{~\nu} \rangle_{ren}$ in the exterior region, excluding the horizon, of a lukewarm black hole. As described in the previous chapter, the radial equation cannot, in general, be solved in closed form for most space-times of interest. While it may be solved in special cases,  for example the $n=0$ mode  radial equation in Schwarzschild space-time, when the quantum field is massless, reduces to Legendre's equation \cite{Candelas};  in general, however, one has to resort to numerical integration to find the desired solutions.
  We begin by introducing the lukewarm black hole space-time, we then consider the numerical integration of the radial equation, before moving on the details of calculating the relevant mode sums.

\subsection{Lukewarm Black Holes}
\label{sec:LW}
 Lukewarm black holes are a special class of Reissner-Nordstrom-de Sitter space-times with (Euclidean) line element given by Eq.~(\ref{lee}) with metric function
\begin{equation}
\label{metric}
f (r) = 1 -\frac{2M}{r} +\frac{Q^2}{r^2} - \frac{\Lambda r^2}{3},
\end{equation}
where $M$, $Q$ are the mass and charge of the black hole respectively, and $\Lambda$ is the (positive) cosmological constant, with $Q=M$. 
For $4M<\sqrt{3/\Lambda}$ we have three distinct horizons, a black hole event horizon at $r=r_h$, an inner Cauchy horizon at $r=r_-$, and a cosmological horizon at $r=r_c$, where
\begin{subequations}
\begin{align}
r_-&=\frac{1}{2}\sqrt{{3}/{\Lambda}}\left(-1 +\sqrt{1 +4M\sqrt{{\Lambda}/{3}}}  \right).\\
r_h&=\frac{1}{2}\sqrt{{3}/{\Lambda}}\left(1 -\sqrt{1 -4M\sqrt{ {\Lambda}/{3}}}  \right).\\
r_c&=\frac{1}{2}\sqrt{{3}/{\Lambda}}\left(1 +\sqrt{1 -4M\sqrt{ {\Lambda}/{3}}}  \right).
\end{align}
\end{subequations}
The fourth  root of $f$ is negative and hence nonphysical.\\
 While the event horizon is formed by the gravitational potential of the black hole, the cosmological horizon is formed as a result of the expansion of the universe due to the cosmological constant \cite{GibHawk}. An observer located between the two horizons is causally isolated from the region within the event horizon, as well as from the region outside the cosmological horizon.\\
If, as the evidence seems to suggest, the universe possesses a cosmological constant \cite{Riess:1998}, then it is more natural to consider a black hole configuration which is asymptotically de Sitter than one which sits in an asymptotically flat universe. Also given that de Sitter space is awash with radiation \cite{GibHawk}, it seems 
rather natural that a black hole in a de Sitter background would be most comfortable in a final configuration in which its event horizon is at the same temperature as the surrounding bath. This state of affairs is realized in the lukewarm case and so the study of such a  black hole configuration is well motivated. In fact the lukewarm case has attracted much interest recently, as evidenced in \cite{Winstanley:2007,PhiSq, Matyjasek:2011}.\\
We shall confine our attention to a single exterior region $r \in [r_h,r_c]$ which has a regular Euclidean section with topology $S^2 \times S^2$ \cite{Mellor}.

\subsection{Numerical Integration of the Radial Equation}
\label{sec:NumericalCalculations}
In this section we consider the numerical integration of the equation
\begin{align}
\label{eqn:mode1}
 & \frac{dS}{dr}\left(r^2 f  \frac{d S}{dr}\right) -\bigg(\frac{n^2 \kappa_0^2}{f} +\frac{l(l+1)}{r^2} +m^2 +\xi R\bigg) S=0,
\end{align}
whose solutions will have a Wronskian satisfying
\begin{align}
\label{eqn:Wrons}
C_{nl}\left[p_{nl} \frac{dq_{nl}}{dr}-q_{nl} \frac{dp_{nl}}{dr}\right]=- \frac{1}{r^2 f}
\end{align}
for a lukewarm black hole space-time as described in Sec. \ref{sec:LW}. In this case, Eq.~(\ref{eqn:mode1}) has two regular singular points, at the event and cosmological horizons, which, henceforth, we will denote by $r_h$ and $r_c$, respectively. 
We apply the standard method described in Sec. \textrm{V} of \cite{PhiSq} to find the initial conditions for $p_{nl}$ and $q_{nl}$ about $r_h$ and $r_c$ respectively.

We then utilize the \textit{NDSolve} algorithm in Mathematica to perform our integrations, with one important modification.  \textit{NDSolve} makes use of an interpolating function to give a solution which is defined continuously on the range of integration. This interpolating function, however, introduces large errors in the calculation and requires a computationally prohibative high degree of precision in order to gain accurate results. To avoid this issue we force  \textit{NDSolve} to calculate the desired solution on a defined grid. We do this by splitting our integration range into a grid, performing the integration between successive points and only storing the end point values. In particular we solve Eq.~(\ref{eqn:mode1}) on a evenly spaced grid of 1000 points to a precision of 25 decimal places. This procedure allows us to obtain the Wronskian of our solutions $C_{nl}$ to be constant to at least $10^{-21}C_{n,l}$ at all grid points for $0\leq l \leq100$, $0\leq n\leq 7$.

Finally we choose to express all the dimensional quantities $(M,Q,r,m)$ in units of $L=\sqrt{3/\Lambda}$ (where $\Lambda$ is the cosmological constant), which has the dimensions of length.

\subsection{Mode Sum Calculations Strategy}
As was discussed in Section \ref{sec:unren} in order to ensure convergence of  the mode sums over $l$ one must subtract off multiples of the delta function, representing the divergent behaviour of the summands. These subtraction terms however, calculated in Section \ref{sec:wkb}, only render the mode sums $O(1/l^2)$. Therefore we would require an almost computationally intractable number of modes to achieve an accurate result. To circumvent this issue, we subtract off a large $l$ approximation to the summand inside the sum, thereby rendering the sum rapidly convergent. We then add back on the sum of the approximation. Now this sum of course also converges like $1/l^2$, but if we have a closed form expression for the approximation we are able to transform this sum into rapidly convergent integrals which can be calculated with great accuracy.

The approximation which is most commonly used for such calculations  \cite{Anderson:1994hg,Howard:1985yg,Winstanley:2007}  is the WKB approximation, as it is in a relatively simple form. However, as is well known \cite{Anderson:1994hg,Winstanley:2007} the WKB approximation suffers from issues with uniformity near the horizons, and as a result it fails to capture the correct behaviour as these horizons are approached, resulting in a lack of sufficient convergence in the numerics near the horizon.

Therefore, the strategy we adopt is as follows:
\begin{itemize}
\item  For the mode sums required to calculate $\langle \hat{T}^{\mu}_{~\nu} \rangle_{ren}$ we use the WKB approximation for all $n$ modes. 
\item We only use these numerical results up to a distance from each horizon where the convergence in the numerical mode sums is still at a reasonable level.
\item We then match this last numerical point to the exact horizon values calculated in Paper I.
\end{itemize}
In practice we found that the numerical calculations were of sufficient accuracy up to a distance of 1 grid point from each horizon, so we simply took the next grid point on each end of our region to be the horizon values calculated in Paper I.

\subsection{WKB Contribution}
\label{sec:wkb}

We adopt the WKB approach of Howard \cite{Howard:1985yg}, that is, we find an approximation to the non-linear equation that $C_{nl}p_{nl}q_{nl}$ satisfies. Letting $\beta_{nl}(r) =C_{nl}p_{nl}(r)q_{nl}(r)$ this equation takes the form,
\begin{align}
\label{eqn:WKB}
\beta_{nl}=\frac{1}{2\chi}\left[1-\frac{1}{\chi^2}\left(\frac{r^2 f}{\beta_{nl}^{1/2}}\frac{d^2(r^2f\beta_{nl}^{1/2})}{dr^2} -\eta\right)\right]^{-1/2},
\end{align}
where
\begin{align}
\label{eqn:chidef}
\chi=\sqrt{n^2 \kappa^2 r^4 +(l+\tfrac{1}{2})^2r^2 f},\nonumber\\
\eta=(m^2 +\xi R)f r^4-\tfrac{1}{4} f r^2.
\end{align}
We desire that our WKB approximation be valid for large $n$  as well as for large $l$ so that we may cancel the renormalization subtraction terms, which are given in terms of divergent sums over $n$. Hence, we look for a large $\chi$ approximation to $\beta_{nl}$, i.e. a large $l$ fixed $n$ or a large $n$ fixed $l$ approximation. To keep track of orders it is convenient to replace $\chi$ by $\chi/\epsilon$, where $\epsilon$ is an expansion parameter which we will ultimately set to unity at the end of our calculation. We then write
\begin{align}
\label{eqn:betaexpan}
\beta_{nl}=\beta_{0nl}\epsilon +\beta_{1nl}\epsilon^2+...,
\end{align}
and expand Eqn~(\ref{eqn:WKB}) for small $\epsilon$. To balance both sides of this equation to lowest order we must have that 
\begin{align}
\beta_{0nl}=\frac{1}{2\chi}.
\end{align}
Substitution of this expression into the coefficient of $\epsilon^2$ yields an expression for $\beta_{1nl}$, which in turn gives an expression for $\beta_{2nl}$ and so on.
We may then express each $\beta_{inl}$ coefficient in the expansion (\ref{eqn:betaexpan}) in the form
\begin{equation}
\label{eqn:WKBexpan}
\beta_{inl}=\sum_{j=0}^{2i} \frac{A_{i,j}}{\chi^{2i +2j +1}}.
\end{equation}
The $A_{ij}$ for the first 2 orders are given in Table \ref{tab:WKBexpancof}, the higher order coefficients have quite long expressions but are easily calculable.

\begin{table}[htb]\centering
\begin{tabular}{ |l|l| }\hline
  &\\
  $A_{0,0}$ & $\frac{1}{2}$ \\
&  \\
   $A_{1,0}$ & $\frac{1}{64} \left(r^4 f'^2-4 r^3 f \left(r f''+3 f'\right)-4 r^2 f^2-16 \eta \right)$ \\
   &  \\
  $A_{1,1}$ &$\frac{1}{32} r^6 n^2\kappa^2 \left(-3 r^2 f'^2+2 r f \left(r f''+4 f'\right)-8 f^2\right)$\\
  &  \\
  $A_{1,2}$ &  $\frac{5}{64} r^{10} n^4\kappa^4 \left(r f'-2 f\right)^2$ \\ 
  &\\\hline
 \end{tabular}
\caption{\label{tab:WKBexpancof}WKB expansion coefficients for a spherically symmetric space-time}
\end{table}
In order to calculate the mode sums required to calculate the stress tensor, we also require approximations to the products
\begin{align}
\label{eqn:der}
 C_{nl}p_{nl} \frac{dq_{nl}}{dr}; \quad C_{nl}\frac{dp_{nl}}{dr} \frac{dq_{nl}}{dr}.
\end{align}
We can obtain the first of these expressions by exploiting the Wronskian condition on $p_{nl}$ and $q_{nl}$ 
\begin{align}
C_{nl}\left[ p_{nl} \frac{dq_{nl}}{dr}- q_{nl} \frac{dp_{nl}}{dr}\right]=-\frac{1}{r^2 f},
 \end{align}
 and the definition of $\beta_{nl}$ to give the identity
 \begin{align}
  C_{nl} p_{nl} \frac{dq_{nl}}{dr}= \frac{1}{2}\left(\frac{d \beta_{nl}}{dr} -\frac{1}{r^2 f}\right).
 \end{align}
  The second expression can be obtained by consideration of 
  \begin{align}
 \frac{d^2 \beta_{nl}}{dr^2}= 2 C_{nl}\frac{dp_{nl}}{dr}\frac{dq_{nl}}{dr}+C_{nl}p_{nl}\frac{d^2q_{nl}}{dr^2}+C_{nl}q_{nl}\frac{d^2p_{nl}}{dr^2}.
  \end{align}
The radial equation allows us to replace the second derivatives of $p_{nl}$ and $q_{nl}$ and together with the definition of $\beta_{nl}$ we obtain the following identity:
    \begin{align}
 C_{nl} \frac{dp_{nl}}{dr} \frac{dq_{nl}}{dr}=\frac{1}{2}  \frac{d^2 \beta_{nl}}{dr^2} +\frac{1}{2}\left(\frac{2}{r} +\frac{f'}{f}\right) \frac{d \beta_{nl}}{dr} \nonumber\\
  -\left(\frac{n^2 \kappa^2}{f^2} +\frac{l(l+1)}{r^2 f} +\frac{m^2 +\xi R}{f}\right)\beta_{nl}.
    \end{align}
 We may now obtain approximations to the quantities (\ref{eqn:der}) by inserting our expansion for $\beta_{nl}$,  to the required order, into the above identities. The error in the numerical integration of the radial equation increases with increasing $l$ and, most severely,  with increasing $n$; above $n=7$ the error in the consistency  of $C_{n,l}$ increases rapidly. If we choose to subtract the WKB approximation to the order which renders the sums over $l$ to converge like $l^{-8}$ and those over $n$ like $n^{-7}$, we need only sum up to a maximum of $l=100$ and $n=7$ to give an answer accurate to $10^{-6}$. To obtain this convergence we require at most a fourth order WKB approximation. (For many of the sums a third order approximation suffices, only when we have an $l^3$ or $n^3$ factor is the fourth order approximation necessary.)
 We may also improve upon this accuracy by using the Levin $u$ transform (see \cite{Levrev} for a detailed discussion of this transform), to speed up the convergence of the sums further.
 We now turn our attention to the calculation of the sums over $n$ and $l$ of the WKB approximation. We will use the transformation
    \begin{align}
    \label{eqn:WS}
  \sum_{l=0}^{\infty}F(l)= \int_{0}^{\infty} F(\lambda-\tfrac{1}{2}) d\lambda- \mathcal{R}\left[ \int_{0}^{\infty}\frac{2F(i\lambda-\tfrac{1}{2})}{1+e^{2\pi \lambda}} d\lambda\right]\nonumber\\ 
  +\mathcal{I}\left[\mathcal{P} \int_{-\tfrac{1}{2}}^{\infty}  F(l) \cot(\pi l)\right], \end{align}
  to perform the sum over $l$. Here $\lambda= l +1/2$, $ \mathcal{R}$,  $\mathcal{I}$ and  $\mathcal{P}$ denote the real, imaginary and principal parts respectively. We note here that Eq.~(\ref{eqn:WS}) comprises of the standard Wastson Sommerfield formula together with an extra term which only contributes if $F(l)$ takes on complex values for real $l$.
  
  To perform the required sums over $l$ of the WKB approximation we note that, in virtue of Eqn (\ref{eqn:WKBexpan}), each sum takes one of the following forms:
  \begin{align}
    \label{eqn:sumllin}
\sum_{l=0}^{\infty} \frac{2 \lambda g_{i}(r,n)}{(n^2\kappa^2r^4 +\lambda^2 r^2 f)^{\tfrac{1}{2} +i}}  -S_{g},\end{align}
  \begin{align}
  \label{eqn:sumlcube}
  \sum_{l=0}^{\infty} \frac{2 \lambda(\lambda^2-\tfrac{1}{4}) h_{i}(r,n)}{(n^2\kappa^2r^4 +\lambda^2 r^2 f)^{\tfrac{1}{2} +i}}  -S_{h},
\end{align}
where $S_{g}$ and $S_{h}$ represent the subtraction terms which are required to make each sum converge.
We will consider sums of the first type; the same argument applies to those of the second form Eqn (\ref{eqn:sumlcube}) above.
Using Eq. (\ref{eqn:WS}) we can write
 \begin{align}
 \label{eqn:realim}
&\sum_{l=0}^{\infty} \frac{2 \lambda g_{i}(r,n)}{(n^2\kappa^2r^4 +\lambda^2 r^2 f)^{\tfrac{1}{2} +i}} -S_g =\nonumber\\
& \int_{0}^{\infty}\left(\frac{2 \lambda g_{i}(r,n)}{(n^2\kappa^2r^4 +\lambda^2 r^2 f)^{\tfrac{1}{2} +i}}-S_{g} \right)d\lambda\nonumber\\ 
&+\mathcal{R}\left[\int^{\infty}_{0} \frac{4 \lambda }{1+e^{2\pi \lambda}}\frac{g_{i}(r,n)}{(n^2\kappa^2r^4 -\lambda^2 r^2 f)^{\tfrac{1}{2} +i}}d\lambda\right].
\end{align}
In the terminology of Eqn (\ref{eqn:WS}), the function $F(l)$ for this case is always real for real values of $l$ so we may drop the extra term. Note that we drop the contribution of the subtraction term to the second integral above as it would be purely imaginary.

We consider the real integral first. Introducing the quantity $a=n\kappa r/f^{\tfrac{1}{2}}$ and changing variable from $\lambda$ to $q=\lambda/a$ we may recast the real integral in the form
 \begin{align}
 \label{eqn:Real}
I_{R}=\frac{a }{r f^{\tfrac{1}{2}}}\int_{0}^{\infty}\left(\frac{2q g_{i} }{(r^2 f a^2)^{i}(1 +q^2)^{\tfrac{1}{2} +i}} -r f^{1/2}S_g \right)dq.
\end{align}
Performing the integration gives
 \begin{align}
I_{R}=\frac{a }{r f^{\tfrac{1}{2}}}\left[\lim_{q\to\infty}\left(2 g_{i}\frac{(1 +q^2)^{\tfrac{1}{2} -i}}{(r^2 f a^2)^{i} (1-2i)} -q r f^{1/2}S_g \right)\right.\nonumber\\
\left.-\frac{2}{(r^2 f a^2)^{i} (1-2i)}g_i\right].
\end{align}
Now it is clear from the above expression that in order for the integral to be finite we require the subtraction term to satisfy the relation,
 \begin{align}
 \label{eqn:Sub}
S_g = \begin{cases} \displaystyle{ \frac{2g_0}{r f^{1/2}}} &i=0. \\0 & i>0. \end{cases}
 \end{align}
 Correspondingly for sums of the form Eq. (\ref{eqn:sumlcube}),  the above relation takes the form
    \begin{align}
     \label{eqn:Subh}
    S_h = \begin{cases}\displaystyle{  \frac{h_0}{r f^{1/2}}\left(2a^2 q^2 -a^2 -\tfrac{1}{2}\right)} &i= 0.\\
    &\\
    \displaystyle{ \frac{2 h_1}{r ^2f} }& i=1.\\&\\0&i>1. \end{cases}
 \end{align}
  Finally we arrive at the result
  \begin{align}
I_{R}=\frac{a^{1-2i} }{(r f^{1/2})^{1+2i}}\frac{2 g_i}{(2i-1)}.
\end{align}
Using Eqns. (\ref{eqn:Sub})  and  (\ref{eqn:Subh}) one can deduce the form of the large $l$ subtraction terms required for each sum in  $\langle \hat{T}^{\mu}_{~\nu} \rangle_{ren}$, which we denote by $S_G$, $S_{Grr'}$ and so forth. These are given in Table \ref{tab:subterms}.
 \begin{table}[!htb]\centering
\begin{tabular}{ |l|l| }\hline
  &\\
  $S_{G
 }$ &$  \displaystyle{\frac{1}{r f^{1/2}}}$ \\
  &\\
  $S_{Gtt'
 }$ &$ \displaystyle{ \frac{1}{r f^{1/2}}}$ \\
&  \\
   $S_{Gtt}$ & $-\displaystyle{\frac{1}{4 r^2 f^{3/2}}}\left(r f'+4 \lambda  \sqrt{f}+2 f\right)$ \\
   &  \\
  $S_{Grr'}$ &$\displaystyle{\frac{1}{32 r^3 f^{5/2}}}\bigg(r^2 \left(3 f'^2-16 n^2 \kappa ^2\right)-4 f (r^2 (1-4 \xi )
   f''$\\
  &$ +(r-16 r \xi ) f'+4 m^2 r^2+8 \lambda ^2+8 \xi -1)+4 (8 \xi +1)
   f^2\bigg)$\\
  &  \\
  $S_{G\theta\theta'}$ &  $\displaystyle{\frac{1}{32 r
   f^{3/2}}}\bigg(r^2 \left(f'^2-16 n^2 \kappa ^2\right)-4 f (r^2 (1-4 \xi ) f''$\\
   & $+r
   (3-16 \xi ) f'+4 m^2 r^2-8 \lambda ^2+8 \xi +1)+4 (8 \xi -1) f^2\bigg)$ \\ 
  &\\\hline
 \end{tabular}
\caption{\label{tab:subterms}Large $l$ subtraction terms}
\end{table}

It is now incumbent upon us to calculate the second integral in Eq.~(\ref{eqn:realim}),
   \begin{align}
I_{C}=\mathcal{R}\left[\int^{\infty}_{0} \frac{4 \lambda }{1+e^{2\pi \lambda}}\frac{g_{i}(r,n)}{(n^2\kappa^2r^4 -\lambda^2 r^2 f)^{1/2+i}}d\lambda\right].
\end{align}
Once more, changing independent variable from $\lambda$ to $q$ allows us to write this integral in the form
   \begin{align}
   \label{eqn:Comp}
I_{C}=\mathcal{R}\left[\frac{a }{r f^{\tfrac{1}{2}}}\int^{\infty}_{0}\frac{\hat{g}_{i}}{(1-q)^{1/2 +i} }dq\right].
\end{align}
For sums of the type Eq. (\ref{eqn:sumlcube}) we obtain an equivalent expression as above with $\hat{g}$ replaced by  $\hat{h}$ 
where
 \begin{align}
\hat{g}_{i}= \frac{4 a q }{1+e^{2\pi aq}}\frac{g_{i}}{(r^2 f a^2)^{i} (1+q)^{1/2 +i}}.
\end{align}
 \begin{align}
 \label{eqn:hh}
\hat{h}_{i}= -\frac{4 aq(a^2 q^2 +\tfrac{1}{4} )}{1+e^{2\pi aq}}\frac{h_{i}}{(r^2 f a^2)^{i} (1+q)^{1/2 +i}}.
\end{align}
Inspection of Eq.~(\ref{eqn:Comp}) enables us to conclude that the integral possesses branch points at $q=\pm 1$, so we must introduce a cut along the plane between $[-1,1]$. 
 \begin{figure}[!htb]\centering
\includegraphics[width=8cm]{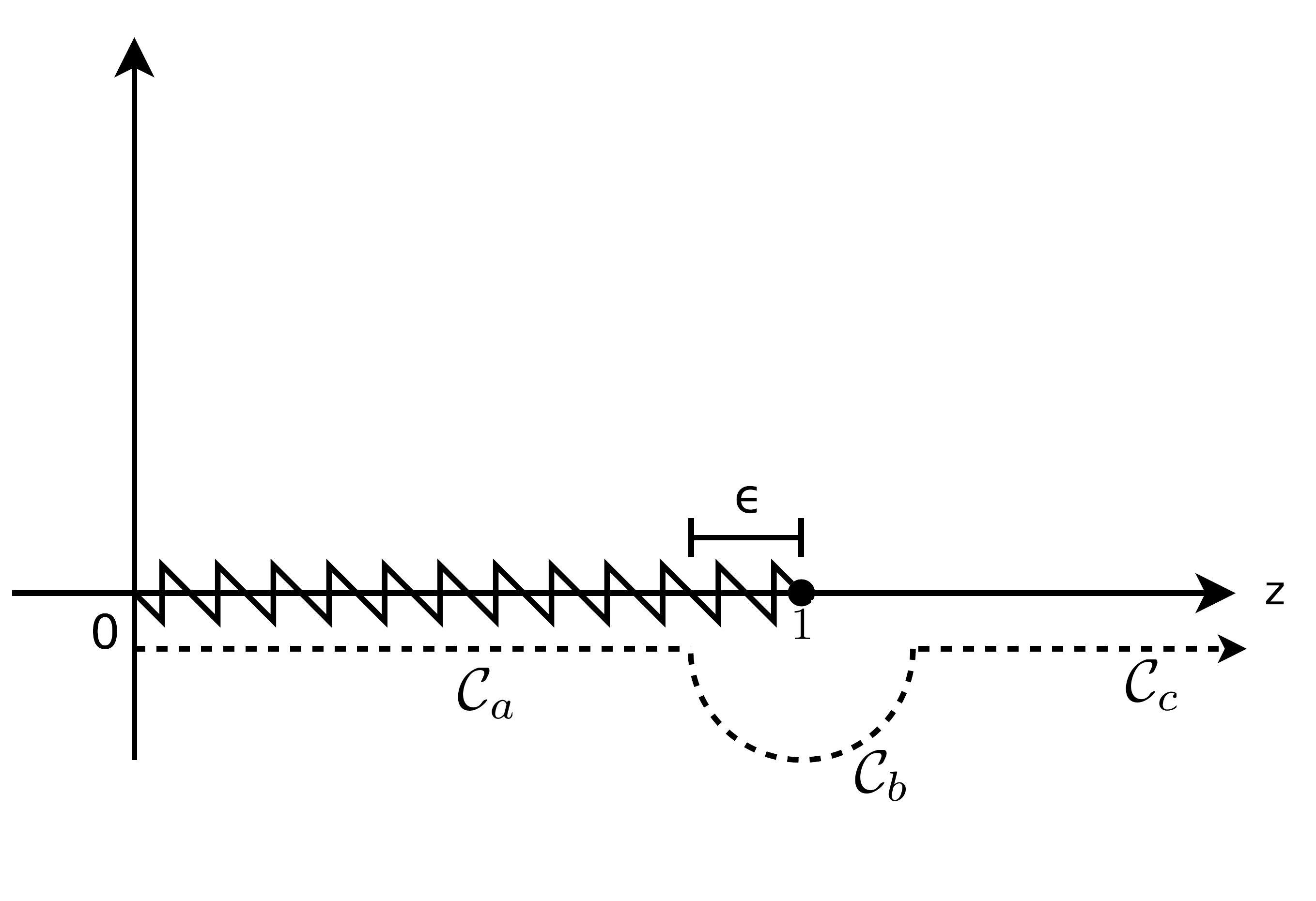}\caption{\label{fig:comcontour} The contour of integration for $I_C$.}
\end{figure}

We now split the integral into 3 integrals along the contours $\mathcal{C}_a$, $\mathcal{C}_b$ and  $\mathcal{C}_c$, shown in Fig. \ref{fig:comcontour}. Thus our variable $q$ to runs from $0 \to 1 -\epsilon$ on $\mathcal{C}_a$ and from $1 +\epsilon \to \infty $ on $\mathcal{C}_c$. It is clear that the integral along $\mathcal{C}_c$ is purely imaginary and so does not contribute to the final answer. Both the integral along $\mathcal{C}_a$  and along $\mathcal{C}_b$ will have a contribution which will be divergent in the $\epsilon \to 0$ limit. In Appendix \ref{ap:WKB} we show that these contributions cancel out leaving a answer which is finite in the $\epsilon \to 0$ limit, given by:
  \begin{align}
     \label{eqn:Ic}
       I_C=& \frac{a }{r f^{\tfrac{1}{2}}}\left(\sum_{j=0}^{i-1} \frac{(-1)^{j} \hat{g}_{i}^{(j)}(1)}{ j! (j-i +\tfrac{1}{2})}\right.\nonumber\\
        +&\left.\int^{1}_{0}\frac{ \hat{g}_{i}(q)- \sum_{j=0}^{i-1} \frac{(-1)^{j}}{j!} \hat{g}_{i}^{(j)} (1)(1-q)^j}{(1-q)^{i +\tfrac{1}{2}}}\right).
        \end{align}
        This result is equally valid if $\hat{g}$ is replaced by $\hat{h}$ defined by Eq.~(\ref{eqn:hh}).
        
       Now while Eq. (\ref{eqn:Ic}) is a perfectly finite expression, in practice we must calculate the integral numerically, which will run into accuracy errors near $q=1$ despite being finite there. This is due to the divergent nature of the denominator in the integral, which for large $i$ will diverge rapidly, so that inevitable round-off errors in the denominator will lead to potentially large errors in the numerical integration. To avoid these errors we integrate by parts $i-1$ times (see Appendix  \ref{ap:WKB} for details) leading to the following expression for $I_C$:
          \begin{align}
          \label{eqnwkbint2}
              I_C= \frac{a }{r f^{\tfrac{1}{2}}} \left[-\frac{g_i(0)}{(i-\tfrac{1}{2})} +\frac{g_i'(0)}{(i-\tfrac{1}{2})(i-\tfrac{3}{2})}+...\right.\nonumber\\
              \left. +(-1)^{i-1}\frac{g_{i}^{i}(0)}{(i-\tfrac{1}{2})(i-\tfrac{3}{2})....(\tfrac{1}{2})}  \right.\nonumber\\
 \left.  +2\frac{(-1)^{i}}{(i-\tfrac{1}{2})(i-\tfrac{3}{2})....(\tfrac{1}{2})}\left( g_i^{(i)}(0) \right.\right.\nonumber\\
 \left. \left.+ \int_{0}^{1}g_i^{(i+1)}(q)(1-q)^{1/2}\right)\right],
        \end{align} 
        which may now be easily calculated numerically.
       Again, the corresponding result, when $\hat{g}$ is replaced by $\hat{h}$,  is also valid. Finally we note that for the $n=0$ mode, from Eq.~(\ref{eqn:chidef}), $\chi= (l+1/2)\sqrt{r^2 f}$, therefore each sum over $l$ is of the simple form:
\begin{align}
H(r)\sum_{l=0}^{\infty}(l+1/2)^{-2j}.
\end{align}
where $j$ is an integer $>0$ and $H(r)$ is some function of $r$. Due to their simplicity, these sums may be directly evaluated with out recourse to the Watson-Sommerfeld method, using \cite{DLMF, gradriz}
\begin{align}
\sum_{l=0}^{\infty}(l+1/2)^{-2j}&= -(1-2^{2n})\zeta(2n)\nonumber\\
&= -\frac{(1-2^{2n})2^{2n-1}}{(-1)^{n-1} (2n!)} B_{2n}.
\end{align}
where $\zeta$ is Riemann's Zeta function and $B_{2n}$ are the Bernoulli numbers.
  
At this juncture it is useful to take stock of our progress thus far. We are now in a position to calculate the sum over $l$ and $n$ of the full radial solutions minus their WKB approximations to a reasonable accuracy ($\approx 10^{-6}$). We may also perform the sum over $l$ of the WKB approximation using the integration method outlined above. Therefore all that remains for us to calculate in this section, is the sum over $n$ of these WKB integrals. We use the Levin $u$ transform to perform this sum over $n$. We calculate the first 20 $n$ values and then apply the Levin u transform to these values to give a final value for the sum. 

We now have all tools required to calculate both $\langle \phi^2\rangle_{ren}$ and $\langle \hat{T}^{\mu}_{~\nu} \rangle_{ren}$ numerically in the exterior region, excluding the immediate vicinity of the horizons. Plots of the results combined with the exact hoirzon values calculated in Paper 1, can be found in the next section. 

\section{Results}
\label{sec:Res}
\subsection{Plots of $\langle \hat{T}^{~\mu}_{\nu} \rangle_{ren}$}
\begin{figure}[!htb]\centering
\includegraphics[width=8cm]{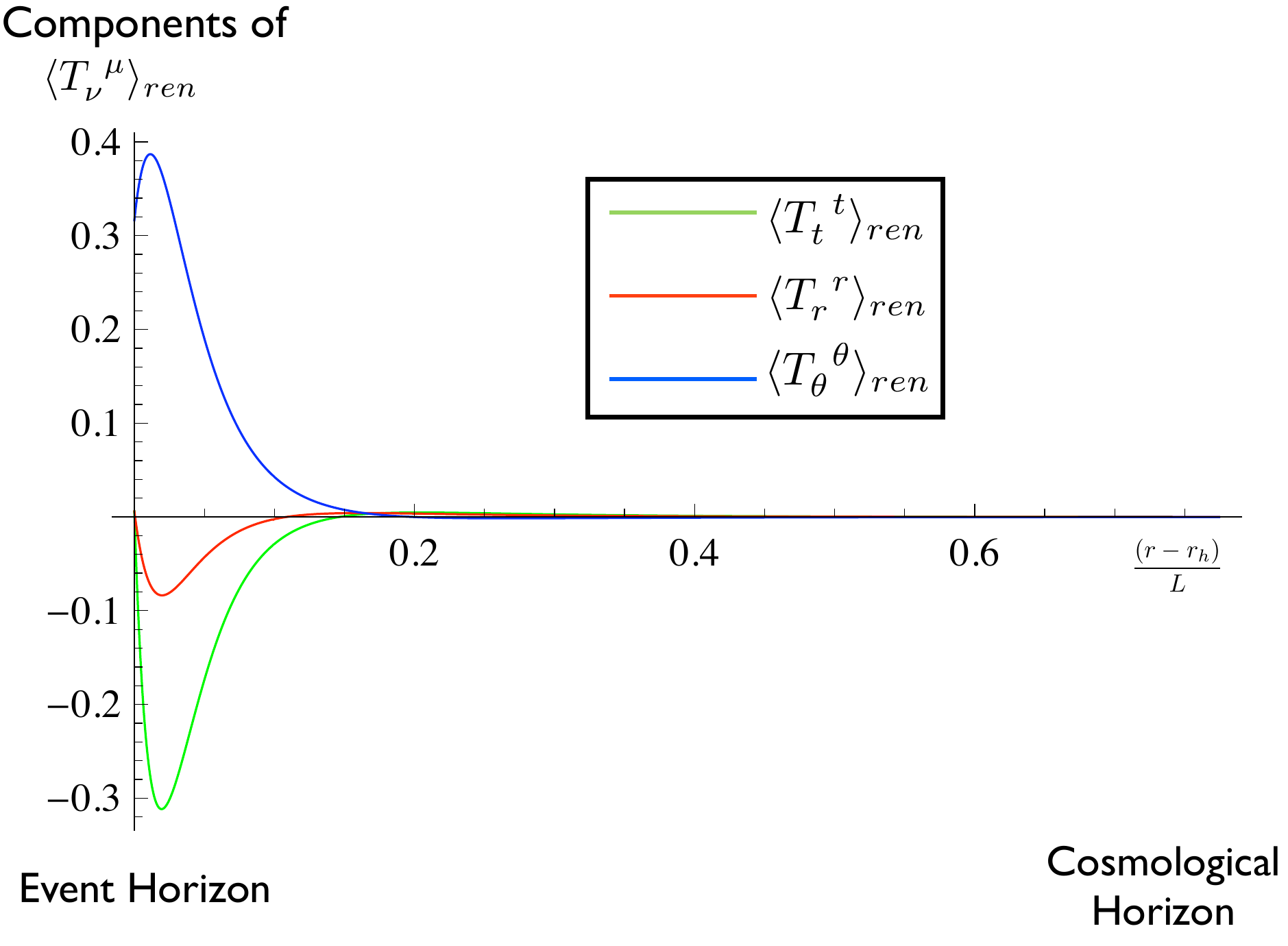}\caption{\label{fig:Tmunurenm0} A plot of the non-zero elements of $\langle \hat{T}^{~\mu}_{\nu} \rangle_{ren}$ in the entire exterior region.}
\end{figure}

\begin{figure}[htb]\centering
\includegraphics[width=8cm]{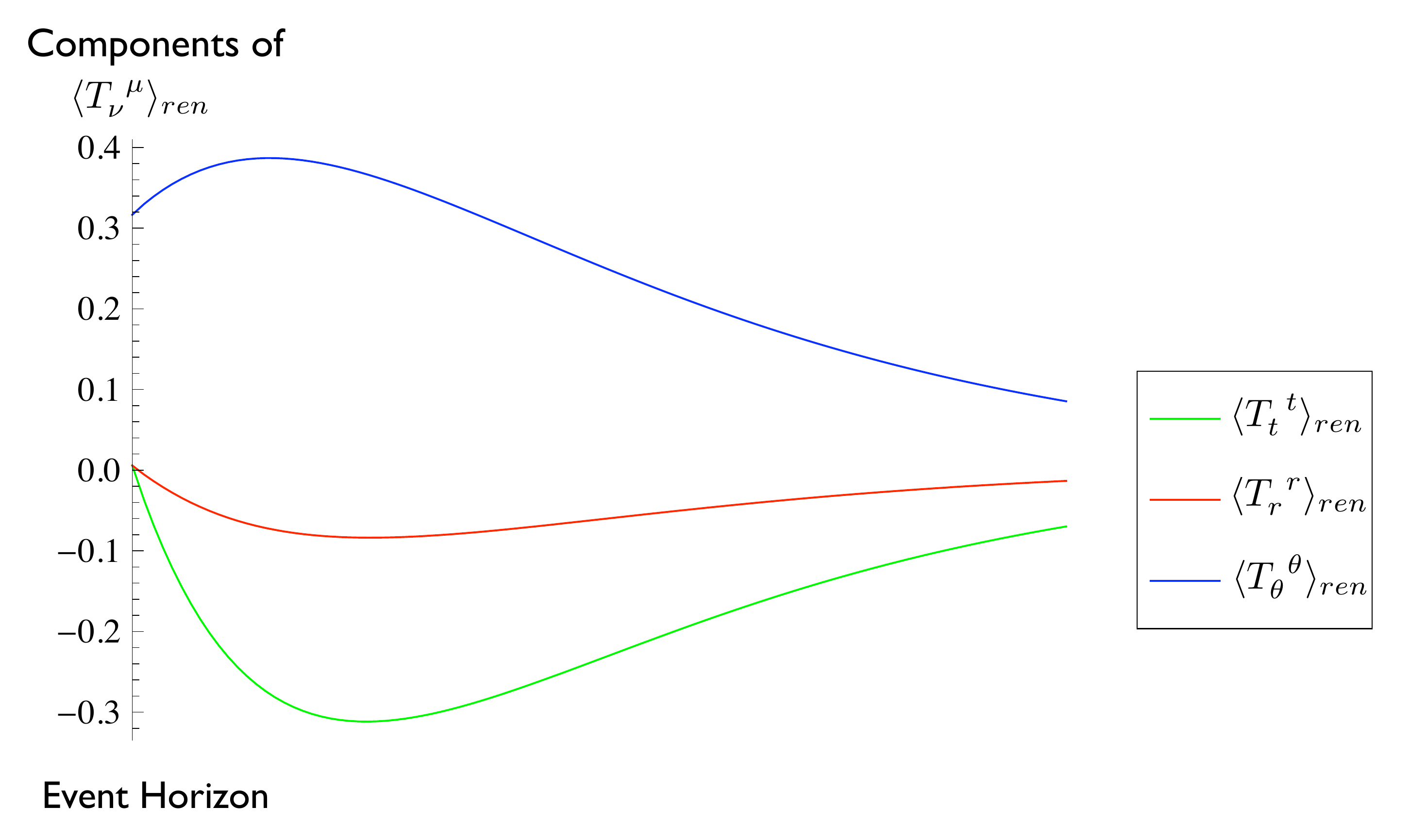}\caption{\label{fig:Tmunum0neh}A plot of the non-zero elements of $\langle \hat{T}^{~\mu}_{\nu} \rangle_{ren}$  in the region of the event horizon.}
\end{figure}

\begin{figure}[!htb]\centering
\includegraphics[width=8cm]{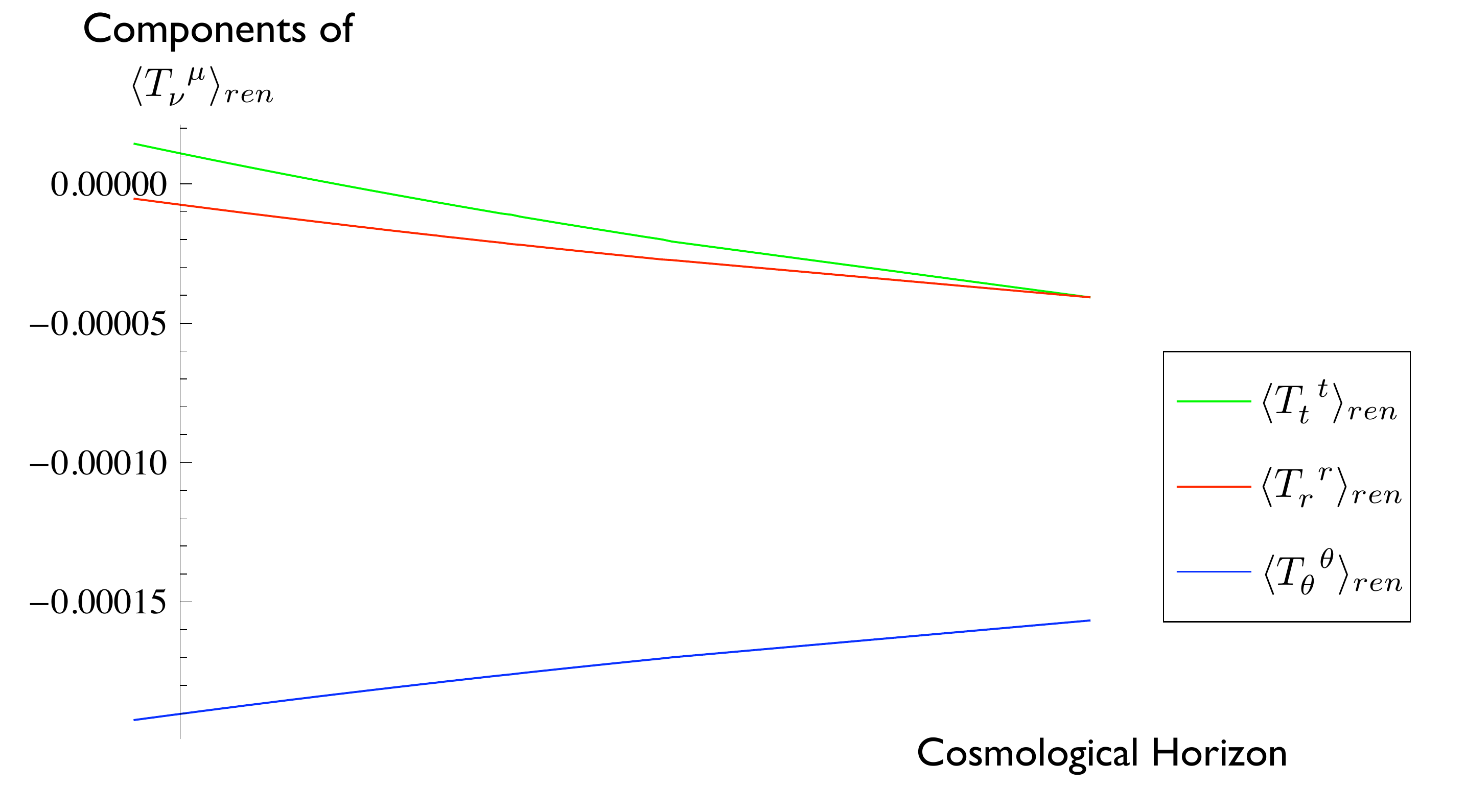}\caption{\label{fig:Tmunum0nch} A plot of the non-zero elements of $\langle \hat{T}^{~\mu}_{\nu} \rangle_{ren}$ in the region of the cosmological horizon.}
\end{figure}

In Fig.~\ref{fig:Tmunurenm0} we plot the components of the stress tensor, calculated on a grid of 1000 points, for a massless conformally coupled scalar field in the exterior region of a lukewarm black hole with $M=Q=0.1L$.
We note here that due to spherical symmetry  $\langle \hat{T}^{~\theta}_{\theta} \rangle_{ren}= \langle \hat{T}^{~\phi}_{\phi} \rangle_{ren}$, so we do not include a plot of the latter.
$\langle \hat{T}^{~\mu}_{\nu} \rangle_{ren}$ is plotted on the first 100 points on the grid plus the event horizon value in Fig.~\ref{fig:Tmunum0neh}, while $\langle \hat{T}^{~\mu}_{\nu} \rangle_{ren}$ is shown on the last 100 viable grid points plus the cosmological horizon value in Fig.~\ref{fig:Tmunum0nch}.
In Figs.~\ref{fig:Tmunum0neh} and \ref{fig:Tmunum0nch}  we see the equality of  $\langle \hat{T}^{~r}_{r}\rangle_{ren}$ and $\langle \hat{T}^{~t}_{t}\rangle_{ren}$ on both horizons as derived in Paper 1.

\subsection{Conservation Equations}

\begin{figure}[!htb]\centering
\includegraphics[width=8cm]{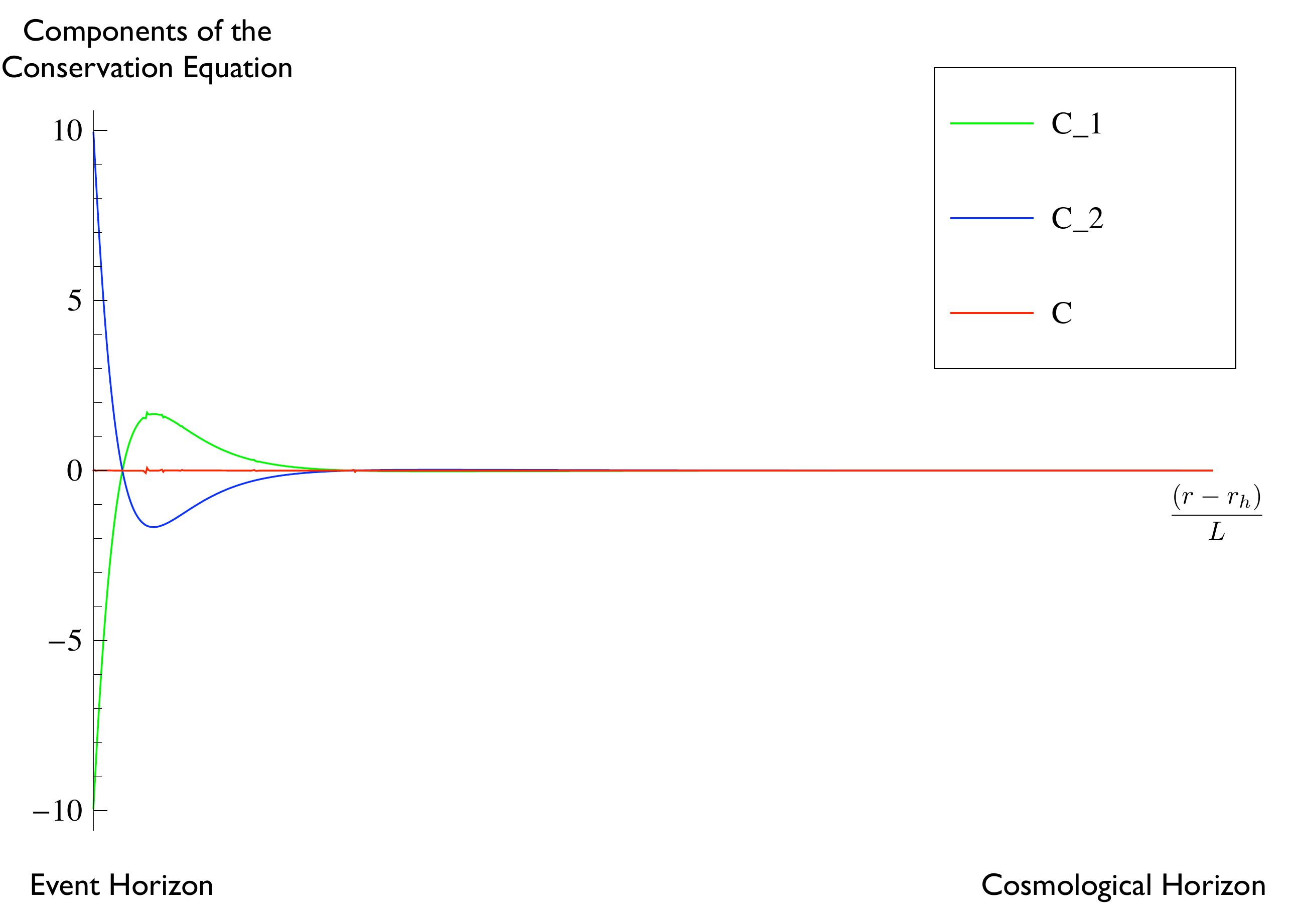}\caption{\label{fig:conserfulllev} A plot of the components of the conservation equation over the entire exterior region.}
\end{figure}

\begin{figure}[!htb]\centering
\includegraphics[width=8cm]{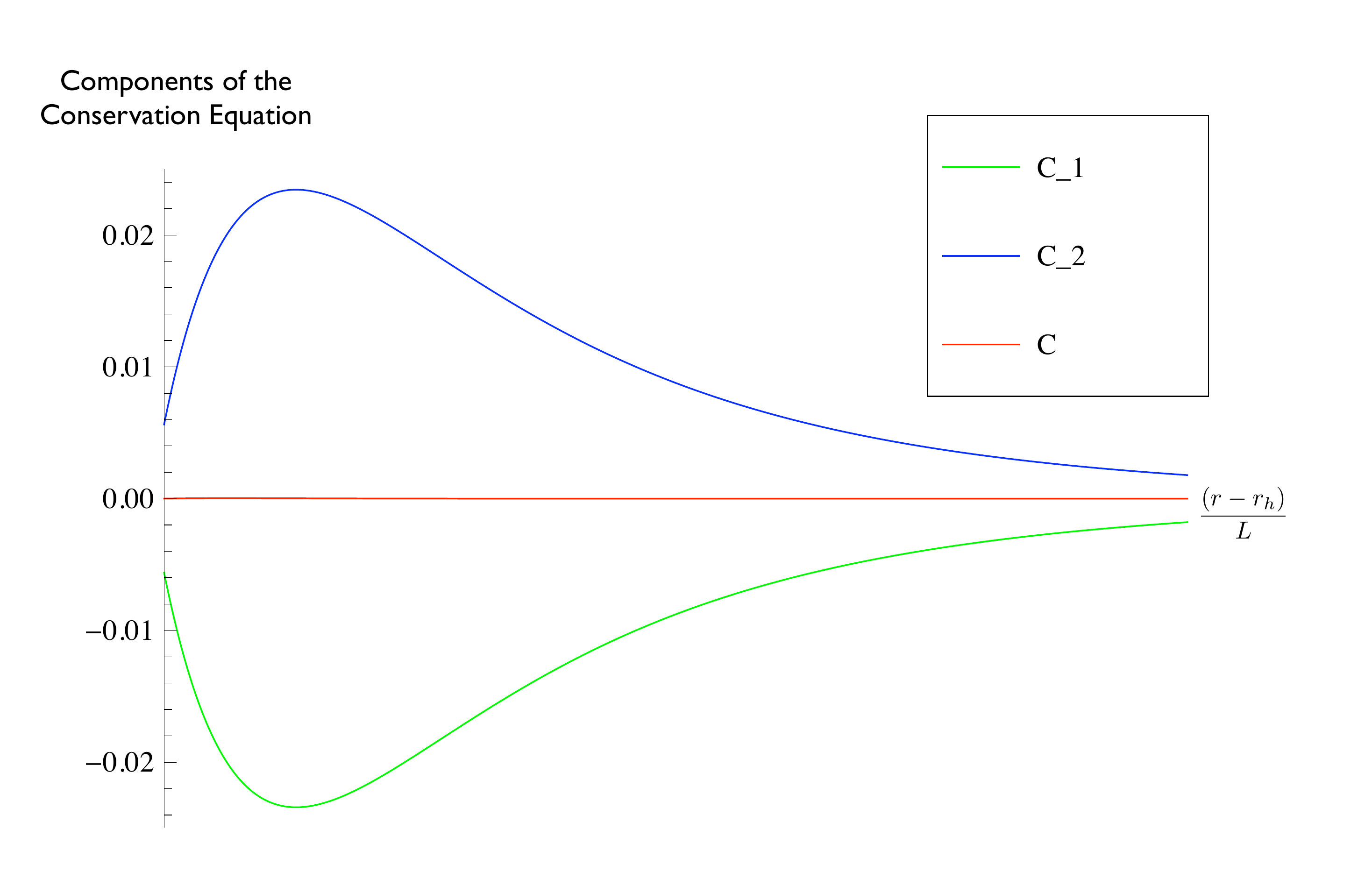}\caption{\label{fig:consermidlev} A plot of the components of the conservation equation over a mid-section of the exterior region.}
\end{figure}

In the Section \ref{sec:conserva} we demonstrated that the analytical contribution to $\langle \hat{T}^{\mu}_{~\nu} \rangle_{ren}$ is a conserved quantity. We are now in a position to perform the same analysis on the numerical contribution for a lukewarm black hole space-time. The conservation equation is given by

\begin{align}
\label{eqn:consernum}
\nabla_{\mu}\langle \hat{T}^{\mu}_{~r} \rangle_{ren}&=\langle \hat{T}^{r}_{~r} \rangle_{ren,r} +\frac{2}{r} \langle \hat{T}^{r}_{~r} \rangle_{ren}\nonumber\\
&+\frac{f'}{2f}(\langle \hat{T}^{r}_{~r} \rangle_{ren}-  \langle \hat{T}^{t}_{~t} \rangle_{ren}) -\frac{2}{r}  \langle \hat{T}^{\theta}_{~\theta} \rangle_{ren}.
\end{align}
As we have performed all of our numerical calculations on a grid we see that in order to construct  the quantity $\langle \hat{T}^{r}_{~r} \rangle_{ren,r}$, we must employ a numerical derivative scheme. We choose to implement a finite difference scheme to calculate the derivative. Mathematica has a inbuilt finite difference algorithm which we employ. Unfortunately, calculating $\langle \hat{T}^{r}_{~r} \rangle_{ren,r}$ using this method is prone to error, in particular near the horizon where the rate of change is large. These errors introduce a lot of numerical noise into the conservation equation. However, a more informative measure, to that of the conservation equation alone, is the magnitude of the conservation equation ($C$) relative to that of its constituent parts, which we denote as
\begin{align}
&C_1=\langle \hat{T}^{r}_{~r} \rangle_{ren,r} ,\nonumber\\
&C_2=\frac{2}{r} \langle \hat{T}^{r}_{~r} \rangle_{ren}+\frac{f'}{2f}(\langle \hat{T}^{r}_{~r} \rangle_{ren}-  \langle \hat{T}^{t}_{~t} \rangle_{ren}) -\frac{2}{r}  \langle \hat{T}^{\theta}_{~\theta} \rangle_{ren}.
\end{align} 
In Fig. \ref{fig:conserfulllev} we see a plot of the conservation equation and its components between the event and cosmological horizon for a massless scalar field on lukewarm black space-time with parameter values $M=Q=0.1L$. In Fig. \ref{fig:consermidlev} we zoom into a mid section of the exterior region, again for a massless scalar field on lukewarm black space-time with the same black hole parameter values.
These plots show that the conservation equation is small relative to its constituent components, $C_1$ and $C_2$, almost everywhere with the exception of the region of the point where $C_1=C_2=0$ and the region of the cosmological horizon. 

In the neighborhood of the cosmological horizon, the magnitude of the quantitates we are calculating approach the error in our numerical scheme, hence the relative error becomes an issue. In fact after the $960$th grid point the values we obtained for the derivatives of the Green's functions are badly behaved. So we ignore these data points and just simply interpolate between the $960$th grid point and the values we will obtain for the derivatives on the horizon in Paper 1.

\subsection{Regularity of the Hartle Hawking State}
\begin{figure}[!htb]\centering
\includegraphics[width=8cm]{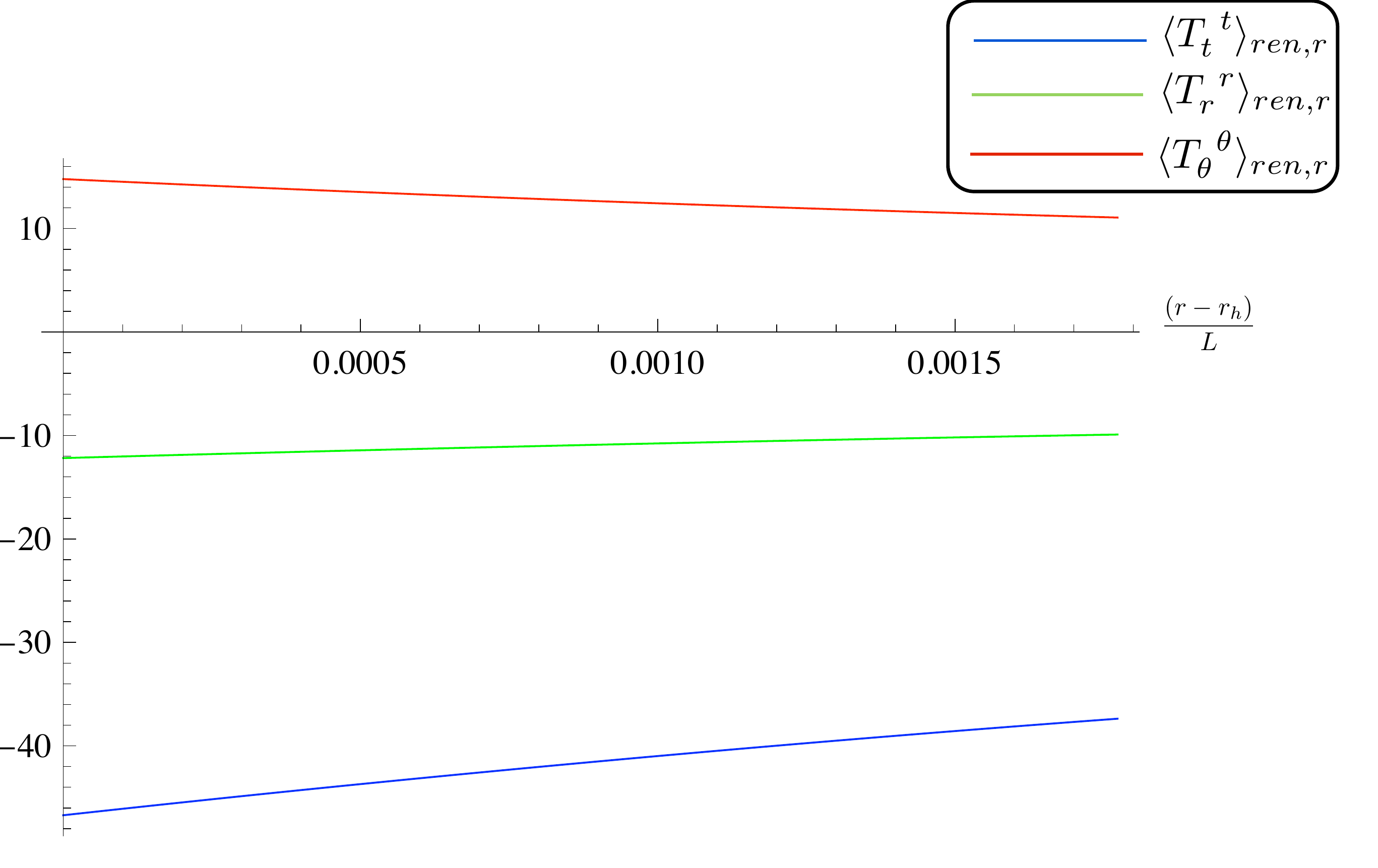}\caption{\label{fig:Tderiv} Radial derivatives of $\langle T_{t}^{~t}\rangle_{ren}$, $\langle T_{\theta}^{~\theta}\rangle_{ren}$ and $\langle T_{r}^{~r}\rangle_{ren}$ in the vicinity of the event horizon}
\end{figure}

In Paper 1 we reduced the constraints on the stress tensor for the regularity of the state to requiring that the components  of  $\langle \hat{T}^{~\mu}_{\nu} \rangle_{ren}$ possesses a Taylor series to the first order about the horizons of the black hole space-time of interest. 
As we have calculated both $\langle T_{r}^{~r}\rangle_{ren}$ and $\langle T_{t}^{~t}\rangle_{ren}$ on the entire exterior region of a lukewarm black hole, we may numerically calculate the radial derivative of each, either by using a finite difference method, or by fitting an interpolating function to the data points and differentiating. In Fig.~\ref{fig:Tderiv} we plot the derivatives of  $\langle T_{r}^{~r}\rangle_{ren}$, $\langle T_{t}^{~t}\rangle_{ren}$ and $\langle T_{\theta}^{~\theta}\rangle_{ren}$  in the region of the event horizon. It is clear that both numerical derivatives are finite there. If this is the case then they must both possess a Taylor series, at least to $O(r-r_0)$, about both horizons. Therefore we may conclude, with corresponding confidence, that the equivalent of the Hartle-Hawking state for a lukewarm black hole configuration, is regular on both the event and cosmological horizons.
   \begin{figure}[!htb]\centering
\includegraphics[width=8cm]{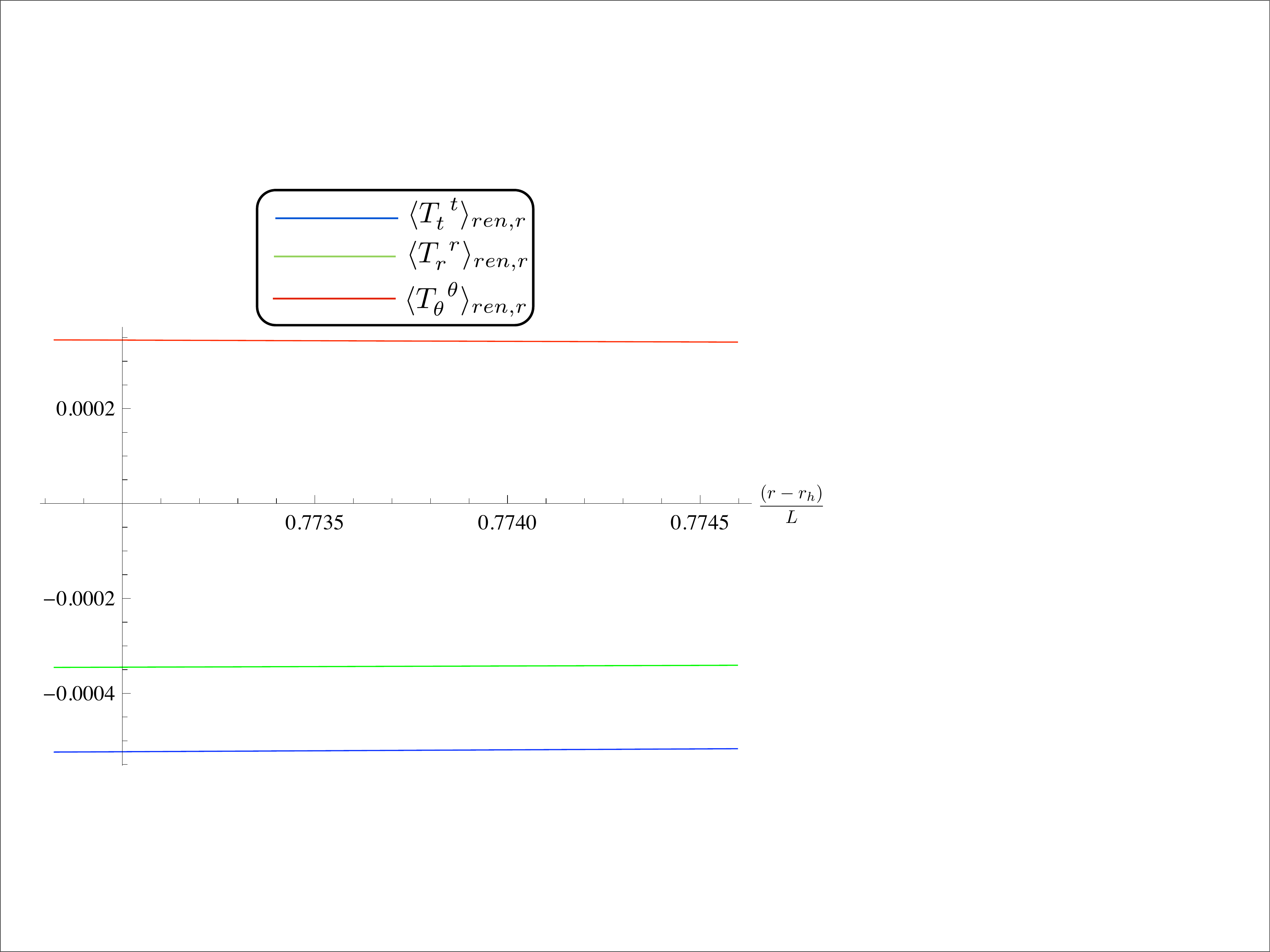}\caption{\label{fig:Tderivc} Radial derivatives of $\langle T_{t}^{~t}\rangle_{ren}$, $\langle T_{\theta}^{~\theta}\rangle_{ren}$ and $\langle T_{r}^{~r}\rangle_{ren}$ in the vicinity of the cosmological horizon}
\end{figure}

\section{Conclusions}
\label{sec:Conclusions}
In this paper, we have, using the Hadamard renormalization procedure, developed a new prescription for the calculation of the  $\langle \hat{T}_{\mu \nu} \rangle_{ren}$ for a spherically symmetric space-time possessing a thermal state. In this new approach, each derivative of the Green's function is renormalized individually, after which they may be combined to form the $\langle \hat{T}_{\mu \nu} \rangle_{ren}$. This method is equivalent to the method of Anderson, Hiscock and Samuel, but has the advantage of being more direct. Our method avoids the separate identification of divergences via the WKB approximation and provides much greater control in debugging numerical calculations.

The application of this new method naturally splits the $\langle \hat{T}_{\mu \nu} \rangle_{ren}$ into two components; an analytic contribution which is formed from a combination of closed form expressions, and a numerical component. We showed that the analytical component is a conserved quantity for a general spherically symmetric space-time. We then calculated, using Mathematica, the numerical component for a field in Hartle-Hawking state on the exterior region (excluding the immediate vicinity of the horizons) of a lukewarm black hole, and demonstrated that it too is conserved. To facilitate this calculation we developed a method of performing the required summations of the WKB approximation of arbitrary order.

Finally combining the results of this paper with the horizon calculations contained in Paper 1, we have plotted the components of  $\langle \hat{T}_{\mu \nu} \rangle_{ren}$  on  the exterior region of a lukewarm black hole. The combination of the results from both papers provides us with compelling numerical evidence that the equivalent of the Hartle-Hawking state for the lukewarm black hole is regular on both the event and cosmological  horizons.

\appendix
\section{Renormalization Subtraction Terms}
\label{ap:Sub}
Here we list the renomalization subtraction terms for temporal separation.
We recall here that if we are dealing with a space-time which has a constant Ricci scalar, such as the lukewarm case, then we can define an effective field mass, $\hat{m}=\sqrt{m^2 +(\xi-1/6)R}$.
\begin{widetext}
\begin{align}
8 \pi^2\{G_{Esing}\}=-\frac{2}{\epsilon ^2 f}+\frac{1}{2} \hat{m}^2 \ln
   \left(-\lambda f \epsilon ^2/2
   \right)-\frac{f''}{12}+\frac{f'^2}{24
   f}-\frac{f'}{6 r},
   \end{align}
\begin{align}
&8\pi^2\{g^{tt'}G_{Esing;tt'}\}=  g^{t t'}\bigg[\frac{12}{\epsilon ^4 f}+\frac{\hat{m}^2}{\epsilon
   ^2}  -\frac{f}{1440 r^4}\left\{r^4 f'' \left(f''+60 \hat{m}^2\right)+4 r^2
   f'^2-2 r^3 f' \left(r f'''+2 f''-60
   \hat{m}^2\right)\right.\nonumber\\
  & \left.-4 r f \left(2 f'+r \left(r^2
   f''''+3 r f'''-f''\right)\right)+4
   f^2-4 \left(45 \hat{m}^4 r^4+1\right)\right\}\ln\left(-\lambda f \epsilon ^2/2\right)\nonumber\\
&+\frac{1}{2880
   r^4 f}\left\{3 r^4 f'^4-12 r^4 f f'^2
   \left(f''+5 \hat{m}^2\right)  +24 r f^3 \left(2 f'+r \left(r^2 f''''+3 r
   f'''-f''\right)\right)\right.\nonumber\\
&  \left.+4 f^2 \left(r^2
   \left(3 r^2 f'' \left(f''-30 \hat{m}^2\right)-4
   f'^2+r f' \left(9 r f'''+28 f''-180
   \hat{m}^2\right)\right)+270 \hat{m}^4 r^4+6\right)
+24 f^4\right\}\bigg],
 \end{align}
    \begin{align}
&8\pi^2\{g^{tr'}G_{Esing;tr'}\}= g^{tr'}\frac{2 f'}{\epsilon ^3
   f^2}, \quad 8\pi^2 \{g^{rt'}G_{Esing;rt'}\}= - g^{rt'}
\frac{2 f'}{\epsilon ^3
   f^2}.
 \end{align}
    \begin{align}
&8\pi^2\{g^{rr'}G_{Esing;rr'}\}= g^{rr'}\bigg[\frac{4}{\epsilon ^4
   f^3}+\frac{r^2 f f''-2 r^2 f'^2+3 \hat{m}^2 r^2 f}{3
   r^2 \epsilon ^2 f^3} +\frac{1}{1440 r^4 f }\left\{r^4 f'' \left(f''+60 \hat{m}^2\right)+4 r^2
   f'^2 \right.\nonumber\\
 & \left. -2 r^3 f' \left(r f'''+2 f''-60
   \hat{m}^2\right)+4 r f \left(r \left(r
   f'''+f''\right)-2 f'\right)+4 f^2-4
   \left(45 \hat{m}^4 r^4+1\right)\right\}\ln\left(-\lambda f \epsilon ^2/2\right)\nonumber\\
   &+\frac{1}{2880 r^4 f^3}\left\{41 r^4 f'^4-4 r^4 f f'^2 \left(26
   f''+15 m^2\right)+4 f^2 \left(10 r^4 f''
   \left(f''+9 m^2\right)+r^3 f' \left(5 r
   f'''-24 f''+60 m^2\right)\right.\right.\nonumber\\
 &  \left.\left.-90 m^4
   r^4-2\right) -8 r f^3 \left(26 f'+r \left(7
   r^2 f'+15 r f'''-25
   f''\right)\right)+8 f^4\right\}\bigg],
   \end{align}
   \begin{align}
&8\pi^2\{g^{\theta\theta'}G_{Esing;\theta\theta'}\}=\frac{4}{\epsilon ^4 f^2}+\frac{-r^2 f'^2+f \left(r^2 f''+2 r f'+6
   \hat{m}^2 r^2+2\right)-2 f^2}{6 r^2 \epsilon ^2  f^2}  -\frac{\ln\left(-\lambda f \epsilon ^2/2\right)}{1440 r^4}\left\{r^4 f''^2+4 r^2 f'^2\right.\nonumber\\
  &\left.-2 r^3 f' \left(r
   f'''+2 f''+60 \hat{m}^2\right) + 120 \hat{m}^2 r^2-4
   -2 r f \left(4
   f'+r \left(r^2 f'''+2 r f'''-2
   f''+60 \hat{m}^2\right)\right)+4 f^2+180 \hat{m}^4
   r^4\right\}\nonumber\\
    &+\frac{1}{2880 r^4 f^2}\left\{11 r^4 f'^4-8 r^2 f^3 \left(r \left(9
   f'''+r f''''\right)+16 f''\right)-2
   r^2 f f'^2 \left(17 r^2 f''+10 r f'+30
   \hat{m}^2 r^2+10\right)+8 f^4\right.\nonumber\\
&   \left.+4 f^2 \left(r \left(5 r
   f'' \left(r^2 f''+6 \hat{m}^2 r^2+2\right)-13 r
   f'^2.+f' \left(5 r^3 f'''+14 r^2
   f''+120 \hat{m}^2 r^2+20\right)\right)-90 \hat{m}^4
   r^4-2\right)\right\},
   \end{align}
\begin{align}
\{g^{\phi\phi'}G_{Esing;\phi\phi'}\}=
\{g^{\theta\theta'}G_{Esing;\theta\theta'}\},
  \end{align}
\begin{align}
&8\pi^2\{g^{tt}G_{Esing;t t}\}=\frac{12}{\epsilon ^4 f^2}+
\frac{r^2 f'^2+2 \hat{m}^2 r^2 f}{2 r^2 \epsilon ^2
   f^2}+\frac{1}{1440 r^4}\left\{-r^4 f'' \left(f''+60 \hat{m}^2\right)-4 r^2
   f'^2+2 r^3 f' \left(r f'''+2 f''-60
   \hat{m}^2\right)\right.\nonumber\\
&   \left.+4 r f \left(2 f'+r \left(r^2
   f''''+3 r f'''-f''\right)\right)-4
   f^2+180 \hat{m}^4 r^4+4\right\}\ln\left(-\lambda f \epsilon ^2/2\right)\nonumber\\
   &+\frac{1}{2880
   r^4 f^2}\left\{-27 r^4 f'^4+12 r^4 f f'^2 \left(4
   f''+25 \hat{m}^2\right)+4 f^2 \left(3 \left(r^4
   f'' \left(f''-30 \hat{m}^2\right) +24 r f^3
   \left(2 f'+r \left(r^2 f''''+3 r
   f'''-f''\right)\right)\right. \right.\right.\nonumber\\
   &\left.\left.\left.-24 f^4 +90 \hat{m}^4
   r^4+2\right)+ 26 r^2 f'^2-2 r^3 f' \left(3 r
   f'''+f''+90 \hat{m}^2\right)\right)\right\},
  \end{align}

   \begin{align}
&8\pi^2\{g^{\theta\theta}G_{Esing;\theta \theta}\}=   -\frac{4}{\epsilon ^4 f^2}  +   \frac{r^2 f'^2-f \left(r^2 f''-4 r f'+6
   \hat{m}^2 r^2+2\right)+2 f^2}{6 r^2 \epsilon ^2
   f^2}\nonumber\\
   &\frac{1}{1440 r^4}\left\{r^4 f''^2+4 r^2 f'^2-2 r^3 f' \left(r
   f'''+2 f''+60 \hat{m}^2\right)-2 r f \left(4
   f'+r \left(r^2 f''''.+2 r f'''-2
   f''+60 \hat{m}^2\right)\right)\right.\nonumber\\
  & \left.  +4 f^2+180 \hat{m}^4
   r^4+120 \hat{m}^2 r^2-4\right\}\ln\left(-\lambda f \epsilon ^2/2\right)+\frac{1}{2880 r^4 f^2}\left\{-11 r^4 f'^4+2 r^2 f f'^2 \left(17 r^2
   f''-20 r f'+30 \hat{m}^2 r^2+10\right) \right.\nonumber\\
  & \left.+4 f^2
   \left(-5 r^2 f'' \left(r^2 f''+6 \hat{m}^2
   r^2+2\right)+13 r^2 f'^2+r f' \left(-5 r^3
   f'''+16 r^2 f''+60 \hat{m}^2 r^2-20\right)+90
   \hat{m}^4 r^4+2\right) \right.\nonumber\\
   &\left.+8 r f^3 \left(30 f'+r
   \left(r^2 f''''-6 r f'''-14
   f''\right)\right)-8 f^4\right\},
        \end{align}
        \begin{align}
\{g^{\phi\phi}G_{Esing;\phi\phi}\}=
\{g^{\theta\theta}G_{Esing;\theta\theta}\}.
  \end{align}
  \end{widetext}
  Note we choose not to include the expression for $g^{rr}G_{Esing;rr}$ as it will not be required.
\section{WKB Integrals}
\label{ap:WKB}
In this appendix we derive the expressions~(\ref{eqn:Ic}) and (\ref{eqnwkbint2}).

To derive Eq.~(\ref{eqn:Ic}) we firstly we consider the integral along $\mathcal{C}_a$. We may isolate the divergent  behaviour of the  integral by initially subtracting off the Talyor series of $\hat{g}_{i}$ about $q=1$ from $\hat{g}_{i}$ to the order which will render the integrand integrable at $q=1$ and then adding back on the integral of this series over the denominator $(1-q)^{1/2 +i}$. Following this procedure we obtain the following expression for the integral along $\mathcal{C}_a$:
\begin{widetext}
 \begin{align}
 \label{eqn:c_1}
\mathcal{R}\left[\frac{a }{r f^{1/2}}\left(\sum_{j=0}^{i-1} \frac{(-1)^{j}}{j!} \hat{g}_{i}^{(j)} (1)\frac{1- \epsilon^{j-i +1/2}}{j-i +1/2} +\int^{1-\epsilon}_{0}\frac{ \hat{g}_{i}(q)- \sum_{j=0}^{i-1} \frac{(-1)^{j}}{j!} \hat{g}_{i}^{(j)} (1)(1-q)^j}{(1-q)^{i +1/2}}\right)\right].
 \end{align}
 For the integral along $\mathcal{C}_b$ we transform integration variable from $q$ to $\theta$ in the following manner; $q\to1 + \epsilon e^{i\theta}$, then our integral becomes
  \begin{align}
\mathcal{R}\left[\frac{a }{r f^{1/2}} \int_{-\pi}^{0}\frac{\hat{g}_{i}(1 +\epsilon e^{i\theta})i \epsilon  e^{i\theta}}{(-\epsilon e^{i \theta})^{i+1/2}} d\theta\right]
  \end{align}
We may then expand $\hat{g}_{n}(1 +\epsilon e^{i\theta})$ about $\epsilon=0$ to obtain an expression which is now amenable to integration. This procedure gives the following form for the integral along $\mathcal{C}_2$:
 \begin{align}
 \frac{a }{r f^{\tfrac{1}{2}}} \sum_{j=0}^{\infty} \frac{(-1)^{j} \epsilon^{j -i +\tfrac{1}{2}} g_i^{j}(1) }{j! (j-i +\tfrac{1}{2})},
    \end{align}
   which, together with Eqn (\ref{eqn:c_1}), allows us to obtain
    \begin{align}
    \label{eqn:Icep}
    I_C= \lim_{\epsilon \to 0}  \frac{a }{r f^{\tfrac{1}{2}}}\left(\sum_{j=0}^{i-1} \frac{(-1)^{j} \hat{g}_{i}^{(j)}(1)}{ j! (j-i +\tfrac{1}{2})} +\int^{1-\epsilon}_{0}\frac{ \hat{g}_{i}(q)- \sum_{j=0}^{i-1} \frac{(-1)^{j}}{j!} \hat{g}_{i}^{(j)} (1)(1-q)^j}{(1-q)^{i +\tfrac{1}{2}}}+ \sum_{j=i}^{\infty} \frac{(-1)^{j} \epsilon^{j -i +\tfrac{1}{2}} g_i^{j}(1) }{j! (j-i +\tfrac{1}{2})}\right).
        \end{align}
 As each term in Eq.~(\ref{eqn:Icep}) is manifestly finite as $\epsilon \to 0$ we make take this limit leaving us with Eq.~\ref{eqn:Ic}.
 
 For the derivation of Eq.~(\ref{eqnwkbint2}) it is useful to introduce a new function 
        \begin{align}
         G(q) =\hat{g}_{i}(q)- \sum_{j=0}^{i-1} \frac{(-1)^{j}}{j!} \hat{g}_{i}^{(j)} (1)(1-q)^j.
         \end{align}
       Then we have that $G(1)=G'(1)=..=G^{(i-1)}(1)=0$ and so integration by parts yields
    \begin{align}     
    \int_{0}^{1}\frac{G(q)}{(1-q)^{i+\tfrac{1}{2}}} =-\frac{G(0)}{(i-\tfrac{1}{2})} +\frac{G'(0)}{(i-\tfrac{1}{2})(i-\tfrac{3}{2})}+\dots +(-1)^{i}\frac{G^{(i-1)}(0)}{(i-\tfrac{1}{2})(i-\tfrac{3}{2})\dots(\tfrac{1}{2})}
    +\frac{(-1)^{i}}{(i-\tfrac{1}{2})(i-\tfrac{3}{2})\dots(\tfrac{1}{2})}\int_{0}^{1}\frac{G^{(i)}(q)}{(1-q)^{1/2}}
       \end{align}  
       Now using the definition of $G(q)$ and Eq.~(\ref{eqn:Ic}) we have
       \begin{align}
            &  I_C= \frac{a }{r f^{\tfrac{1}{2}}} \left(-\frac{g_i(0)}{(i-\tfrac{1}{2})} +\frac{g_i'(0)}{(i-\tfrac{1}{2})(i-\tfrac{3}{2})}+... +(-1)^{i}\frac{g_{i}^{i}(0)}{(i-\tfrac{1}{2})(i-\tfrac{3}{2})....(\tfrac{1}{2})}\right.\nonumber\\
&\left. +g_i(1)\left[-\frac{1}{i -\tfrac{1}{2}} +\frac{1}{i -\tfrac{1}{2}}\right]\right.\nonumber\\
&\left.  +g'_i(1)\left[\frac{1}{i -\tfrac{3}{2}} -\frac{1}{i -\tfrac{1}{2}}-\frac{1}{(i -\tfrac{1}{2})(i -\tfrac{3}{2})}\right]\right.\nonumber\\
&\left.  +g''_i(1)\left[-\frac{1}{i -\tfrac{5}{2}}\frac{1}{2!} +\frac{1}{i -\tfrac{1}{2}}\frac{1}{2!}+\frac{1}{(i -\tfrac{1}{2})(i -\tfrac{3}{2})}\frac{1}{1!}+\frac{1}{(i -\tfrac{1}{2})(i -\tfrac{3}{2})(i -\tfrac{5}{2})}\right]+ \dots\right.\nonumber\\
&\left. +(-1)^{i-1}g^{i-1}_i(1)\left[-\frac{1}{\tfrac{1}{2}}\frac{1}{(i-1)!} +\frac{1}{i -\tfrac{1}{2}}\frac{1}{(i-1)!}+\dots+\frac{1}{(i -\tfrac{1}{2})(i -\tfrac{3}{2})\dots\tfrac{1}{2}}\right]\right.\nonumber\\
&\left.  +\frac{(-1)^{i}}{(i-\tfrac{1}{2})(i-\tfrac{3}{2})....(\tfrac{1}{2})}\int_{0}^{1}\frac{g_i^{(i)}(q)}{(1-q)^{1/2}}\right) 
\end{align}
All the quantities in square brackets can be shown to vanish. Therefore we arrive at the desired result, namely Eq.~(\ref{eqnwkbint2})
\end{widetext}
\bibliography{database}

\end{document}